\documentclass{article}

\usepackage[english]{babel}

\usepackage[a4paper,top=2cm,bottom=2cm,left=2cm,right=2cm,marginparwidth=1.75cm]{geometry}

\usepackage{float}
\usepackage{amsmath}
\usepackage{graphicx}
\usepackage{wrapfig}
\usepackage{subfigure}
\usepackage{abstract}
\usepackage[utf8]{inputenc}
\DeclareUnicodeCharacter{2731}{\ensuremath{\star}}
\usepackage[colorlinks=true, allcolors=blue]{hyperref}

\usepackage{setspace}
\title{Auditing health-related recommendations in social media: A Case Study of Abortion on YouTube}

\author{Mohammed Lahsaini$^{a}$, Mohamed Lechiakh$^{a}$\thanks{Corresponding author. E-mail address: mohamed.lechiakh@um6p.ma (Mohamed Lechiakh) }, Alexandre Maurer$^{a}$\\
	$^{a}$ \small UM6P College of Computing, Mohammed VI Polytechnic University, Ben Guerir, Morocco  
}

\date{}

\begin{document}

\maketitle

\begin{abstract}
Recommendation algorithms used by social media, like YouTube, significantly shape our information consumption across various domains. Their influence is especially critical in healthcare, where they can endanger lives by promoting misleading health information and practices. Hence, algorithmic auditing becomes crucial to uncover their potential bias and misinformation, particularly in the context of controversial topics like abortion. We introduce a simple yet effective sock puppet auditing approach to investigate how YouTube recommends abortion-related videos to individuals with different backgrounds. First, we generate user data from YouTube recommendation sessions using simulated individual interactions based on different watching histories and opinions about abortion. Second, we employ graph analysis methods to identify the most influential recommendations made to users among 11174 videos. Then, we evaluate our findings on bias and misinformation in YouTube's abortion-related recommendations. Our findings show that the user’s watch history does not substantially affect the type of recommended videos. In addition, YouTube predominantly promotes pro-abortion content, with only 4\% being misleading, as shown by the 58.4\% pro-abortion videos in their prominent recommendations. However, We found that individuals with medical backgrounds or extensive  histories of medical-related content are less likely to come across abortion myth-debunking videos, while those who have engaged with feminism-related material are more likely to encounter anti-abortion videos. We emphasize the black-box evaluation of YouTube recommendations and assess the veracity of its claims that it has removed deceptive content about abortion. Our framework allows for efficient auditing of recommendation systems, regardless of the complexity of the underlying algorithms.


\end{abstract}

\section{Introduction}
\subsection{Background}

Platforms for user-generated videos, like YouTube, have seen a sharp increase in popularity in recent years. For many people, it has emerged as one of the most important information sources \cite{newsReport}. Unfortunately, these platforms can be a source for the dissemination of false information that might harm individuals and alter their viewpoints and decisions \cite{misinfo2}, such as health-related misinformation \cite{healthinfoonyoutube, misinfo1,HoangLouis}. They can also be aligned with some viewpoint on specific issues  \cite{provac} or exhibit undesirable behaviors \cite{beh1, beh2}. Auditing algorithmic recommendations provided by social media platforms \cite{bartley2021auditing} is critical to improving the reliability and trustworthiness of their content, particularly when it comes to health knowledge. 

Because there is a substantial amount of user-created data being shared on these platforms, they must depend on algorithms that provide features like personalization, recommendations, and content filtering. This is the case with YouTube \cite{YTstats}, in which video suggestion and feed personalization, as well as enforcing community guidelines \cite{YTguidlines} (that specify what sort of material is and is not permitted on the network), cannot be achieved by humans alone. However, these algorithms may use probabilistic decision-making techniques that are too complex for humans to comprehend, or too costly to monitor. 

As a result, efforts to audit these algorithmic solutions have grown more common \cite{WhoAudits, algoaudit}, since auditing provides an operational trail that demonstrates bias and promotes transparency.

\subsection{Motivation}
\label{motivation}
Social media platforms may contribute to a safer and more reliable online environment for healthcare content by improving algorithmic control and rigorous auditing, encouraging greater confidence among users seeking accurate and reputable health information. In this context, abortion is a major healthcare concern that faces numerous challenges, as a result of the spread of misinformation and biased opinions on social media platforms. As a very sensitive subject, abortion discussions often become polarized, leading to the spread of inaccurate information and misconceptions. 

On June 24, 2022, the US Supreme Court overruled the historic Roe v. Wade decision, in which the Court determined that the United States Constitution granted the right to abortion. This means that decisions on the legality of abortion are left to the states, making access to abortion vary dramatically across the country \cite{NYT}. This sparked several rallies around the country, pushing the problem into the spotlight. According to Google Trends, there has been an increase in searches for keywords such as ``abortion'', ``abortion in US'', ``women's rights'', ``ban abortion'', ``abortion laws'', ``Roe v Wade'', ``Post Roe'', ``pro-choice'', and ``pro-life''.

Consequently, Google blogged that it will automatically delete location history when users visit sensitive places like abortion clinics \cite{google}. Three weeks later, YouTube announced  its intention to stop the spread of incorrect information and dangerous advice about abortion \cite{tweet,cnn}. This involves removing videos that instruct users on how to perform hazardous abortions or encourage untrue claims.

YouTube's involvement might have an impact beyond the US, since countries differ in their views on abortion, and there is controversy over its legality internationally. This motivates us to audit the platform to see if it has been successful in removing content that spreads misinformation about abortion, and whether YouTube tries to promote one viewpoint, in the anti-/pro-abortion debate, more than another.

\subsection{Objective}

YouTube's algorithms serve two key purposes: search and recommendation. The search engine assists the user in finding videos quickly by suggesting queries based on search terms connected to the user search and watch history, as well as what other users have previously looked for \cite{YTsearch}. It then identifies relevant materials, and ranks them by order of significance. The recommendation algorithm suggests videos that the user may find interesting. These suggestions are displayed as a list of recommendations on the right side of the screen while a video is playing, and as a feed when the user visits the home page or other sections (trending, subscriptions\dots).

YouTube's algorithms bias can be present in the search results \cite{searchbias, searchbias2}, in the homepage feed \cite{homepage, homepage2} or in the list of recommendations (we will refer to it as YouTube recommendations), which this study tries to measure.

While YouTube has not disclosed the technical details of its recommendation algorithm, they are likely to still be using a deep learning process \cite{YTRecSys} in which candidates are first selected from a corpus of videos based on user history and context, and then ranked based on the features of the videos and other factors. However, without a thorough understanding of the algorithm, the whole video database upon which the algorithm operates, and the user's behaviors in response to algorithmic suggestions, the recommendation system (RS) is essentially a “black box'' from the perspective of social scientist, regulators, or external auditors.

For this reason, we study YouTube recommendations empirically. We implement a data gathering mechanism to mimic users watching a series of recommended YouTube videos under various experimental scenarios. The most prominent videos are then classified based on the trustworthiness of their information and their viewpoint on the issue. Finally, we examine the empirical findings to look at how YouTube recommends abortion-related videos.

\section{Literature review}
\subsection{Misinformation and bias in recommendation systems}
There are many potential sources of bias in a recommendation system that generally reflect individuals' own cognitive and societal biases \cite{10.1145/3209581}. These biases can range from individual behavioral biases (selective exposure \cite{doi:10.1080/15205436.2020.1714663}) to data bias (homophilous social network~\cite{doi:10.1126/science.aaa1160}) and algorithmic bias (filter bubble \cite{filterbubble}). Thus, the loop effect resulting from combining all these bias types represents a vicious cycle wherein social recommendation platforms progressively amplify ideological bias. This entails suggesting political content that aligns with users' past engagement, leading to a reinforcing loop of exposure to ideologically similar viewpoints. Misinformation refers to false or inaccurate information that is deliberately propagated to intentionally cause public harm and misleading, or for profit \cite{10.1145/3373464.3373475}. Misinformation and bias are linked dynamics on social media platforms such as YouTube. Given the various methods of reinforcing users' existing biases, this can make users more susceptible to accepting and sharing misinformation that confirms their preconceived ideas \cite{zimmer2019fake,muhammed2022disaster}. For instance, the algorithmic amplification and the virtual filter bubble construction contribute easily to the rapid spread of misinformation within specific YouTube communities, which shows the intricate interplay between personal beliefs, recommendation algorithms, and the distribution of unreliable content. YouTube, a widely used online social media platform, faces significant concerns related to its recommendation system. Its recommendation algorithm has been described as a "long-term addiction machine" \cite{youtubeRadical}, which has been criticized for putting its users in "rabbit holes" \cite{ledwich2019algorithmic}. Given that recommendations drive 70\% of YouTube's content consumption \cite{youtube70}, the potential for unintentionally exposing audiences to more and more biased content becomes notably concerning. 

\subsection{In the context of health-related content}

It is critical to emphasize the importance of misinformation and biased content particularly in the context of health topics and information. The possibility of accidentally exposing users to biased or inaccurate health content is not simply an issue of misinformation promotion, but also a serious threat to public health \cite{sylvia2020we,chou2018addressing}. The impact of recommendation system dynamics on health issues highlights the critical necessity for responsible algorithmic thinking, auditing and content curation to protect the well-being of individuals seeking trustworthy health information.

While the scientific community has mainly focused on the consequences of misinformation about vaccines and communicable diseases in the context of public health (e.g., COVID-19 misinformation \cite{yttrex,Lie008334,10.1145/3485447.3512039}), there is an important gap in addressing individual health-related misinformation. The spread of abortion misinformation is a striking example of this gap \cite{patev2021towards,sharevski2023abortion}. Personal decisions and understandings of abortion-related misinformation could definitely contribute to shaping public knowledge and attitudes, and propagating societal considerations and actions. This is particularly important in this era of big data and giant social media platforms, where prior studies have shown that 70.1\% of women obtain information regarding abortion from the
internet \cite{LITTMAN201419}. Prior works have focused on investigating abortion misinformation in internet and social media through empirical studies that delve into users' experiences with abortion-related content \cite{sharevski2023abortion,patev2021towards,wu2023medication}. These studies mainly identify the various types and key sources of misinformation \cite{sharevski2023abortion,wu2023medication}, provide statistics based on user interaction metrics \cite{wu2023medication}, and examine users' strategies for coping with this misleading information \cite{sharevski2023abortion}. To the best of our knowledge, our work is the first study to present an auditing framework for misinformation and bias of abortion-related recommendations on YouTube.

\subsection{Auditing YouTube recommendations}
Over the past decade, the research community has conducted several audit studies on YouTube, concentrating on diverse methodological approaches to analyze the potential of the platform's recommendation algorithms for promoting radicalization and misinformation \cite{10.1145/3568392,abul2020examining,hussein2020measuring}. An interesting line of research involves using active measurements, often involving sock puppets (without previous watching history), to investigate algorithmic pathways that steer users from ideologically moderated content to more extreme content via video and channel recommendations \cite{beh2,api2}. These works illuminate how YouTube's recommendation system might accidentally guide users toward polarized and extreme content. In the same context, there have been some works \cite{hosseinmardi2021examining,beh2} that used passive measurements to prove the same result based on user interaction histories (e.g., previous watched videos, clicks, comments, ...). Here, the sock puppet training is
crucial for personalized recommendations and observing the loop bias effect in practice \cite{searchbias,10.1007/978-3-030-76228-5_8,10.1145/3479556,Papadamou2020ItIJ}. However, the audit using real user activity cannot clearly separate the effects of recommendation algorithms from other factors driving radicalization and bias amplification. 

Misinformation and biased content in YouTube recommendation have always been a major concern to judge the platform ethical policy and quality of services. YouTube has already announced efforts for mitigating these effects in its recommendations \cite{youtube4resp}. In the literature, there are many intervention strategies that can be used to address this problem. These proposed approaches can essentially be implemented from a system design perspective using pre-processing \cite{10.1145/3397271.3401321,10.1145/3289600.3291002,10.1145/3397271.3401083}, or in-processing \cite{10.1145/3219819.3219886,10.1145/3437963.3441824,10.1145/3219819.3220122}, or post-processing approaches \cite{10.1145/3531146.3533136,10.5555/3295222.3295319}. Indeed, such solutions have the potential to be valuable for auditing recommendations. However, they come with a significant requirement: having in-depth knowledge and special access to the platform's algorithms and models for implementation and modification. In other words, the platform itself would need to actively implement and test these methods internally, which is a complex task and often not feasible. Therefore, auditing black-box systems has always been a challenging topic poorly addressed by the research community. Among the recently published related works about this topic, Srba et al. \cite{10.1145/3568392} used a sock puppet audit methodology that demonstrated the possibility to ``burst the bubbl'' of misinformation videos on YouTube by consuming videos debunking that misinformation. Carrascosa et al. \cite{10.1145/2716281.2836098} also proposed a trained sock puppet approach for measuring and understanding behavioural targeted advertising in the online advertising market. Also, Figueiredo et al. \cite{10.1145/3372923.3404787} explored how targeted YouTube video-ads meet the regulatory policies (e.g., ethics and laws) regarding children advertising in Brazil and Canada, and Le et al. \cite{10.1145/3308558.3313682} developed a sock puppet auditing system to investigate the personalization of web search results  based on users' browsing histories, which can be inferred by search engines via third-party partnerships. In an interesting study on Twitter, Bartley et al. \cite{bartley2021auditing} used a sock-puppet audit and provided evidence that algorithmic curation of content systematically distorts the information people see.

In the same context, we present in this work a trained sock-puppet audit methodology (using artificial user-bots) for abortion-related videos on YouTube, which is audited as a black-box RS. The choice of the abortion topic is motivated by its significant societal and health implications, as well as the problematic content it presents. Moreover, its exploration within the context of auditing RS is novel from a scientific and research perspective. 

\section{Material and methods}
\begin{figure}[h]
  \centering
  \includegraphics[width=\textwidth]{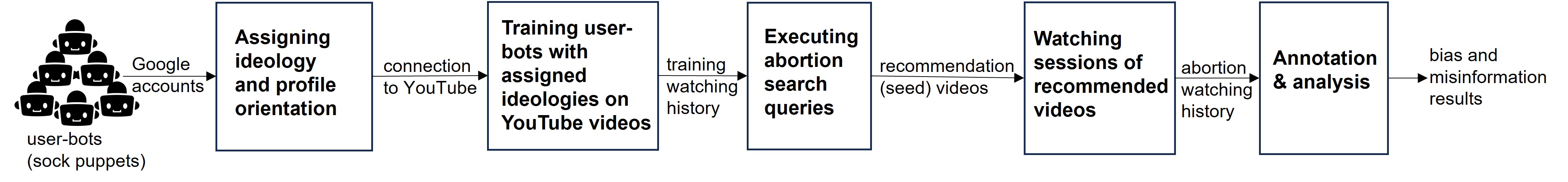}
  \caption{Our sock puppet auditing system to measure bias and the spread of misinformation of abortion-related content on YouTube.}
    \label{audi_model}
\end{figure}

The overarching goal of our study is to examine the extent to which YouTube recommendations spread misinformation and biased content about abortion. Given the difficulty of auditing black-box systems such as YouTube, we aimed to replicate the typical customary browsing and watching behavior of individuals. To this purpose, we take into account the different prevalent ideologies and personal attitudes of individuals seeking abortion-related information online. Thus, we design and implement a sock puppet auditing system to train user profiles through watching sessions of recommended videos on YouTube about abortion, guided by specific search query rules and video selection criteria. Figure \ref{audi_model} shows the general process and steps of our auditing system that are discussed below.

\subsection{Data collection}

Sandvig et al. \cite{Sandvig} reviewed several black box algorithm audit designs, in particular: code audit (when the algorithm is disclosed), noninvasive user audit (which rely on users' answers to questions about how they interact with the platform), scraping audit (in which repeated queries are sent to the platform to observe the outcome), sock puppet audit (where a program assumes the role of a user), and crowdsourced audit (that uses hired users). For our study, we opted for an approach similar to the sock puppet audit, because it offers us a great deal of flexibility in the manipulation of fake users and in data collection.

Using bots, we mimic users who are seeking information about abortion on YouTube. Each bot begins by watching videos gathered from the results of specific search queries on YouTube. While doing so, the recommended abortion videos are collected to be watched next in the following iteration. Each sequence is collected using a distinct Google account with different watch and search history on YouTube. We finally processed over 38k videos, 11174 of which are unique. We detail each stage of the experiment below. The videos have been viewed across multiple sessions over a week to ensure more accurate and relevant recommendations, reduce noise, and align closely with user preferences. This method efficiently minimizes the influence of noise in the final analysis, providing more reliable and meaningful insights.

\subsubsection{Data collection process}

Although YouTube provides an official API that allows the collection of data and has been used in many studies \cite{provac, api1, api2}, we do not use it in our work. Instead, we use Selenium and Python to simulate user behavior and collect recommended video features. The reason for this is that we want to study the effect of the user's watch history on the recommendations, but individual personalizations are not applied to suggestions provided by the API \cite{yttrex}. In fact, they are based on item-to-item similarity and overall user engagement and satisfaction metrics. The consideration of interface affordance shows that users are presented with video recommendations through five distinct pathways: recommendations following keyword queries in the search bar; personalized suggestions based on user interactions; a feed featuring trending videos in specific geographic regions; and  a segment displaying new or unwatched content from subscribed channels. For the scope of this study, our focus will be exclusively directed towards recommendations appearing after querying abortion-related keywords.

\subsubsection{Simulated profiles}

To investigate how YouTube suggestions vary based on what a user has previously seen or looked for, we created six profiles using different YouTube accounts (Google accounts) and had them search for and watch specific types of content on Google and YouTube:
\begin{itemize}
    \item Profile 1: radical feminism and general pro feminism.
    \item Profile 2: anti feminism, male supremacy, incel (Males who may exhibit misogynistic behavior and believe that they are unjustly treated by women and society.).
    \item Profile 3: no search or watch history.
    \item Profile 4: medical lecture and trustworthy medical content.
    \item Profile 5: conspiracy theories (flat earth, UFO\dots).
    \item Profile 6: no search or watch history.
\end{itemize}

In each profile, except for Profile 3 and 6, before we start any collection of data, a bot connects to the appropriate account and executes specific search queries on Google and YouTube, as well as watching videos and visiting websites according to the ideology (theme) of the profile. We ensure that the bot watches 3 minutes of selected videos and stays on each website for at least 5 seconds. According to \cite{homepage}, for YouTube to start offering more customized suggestions, a user must view a minimum of 22 videos. Therefore, the number of videos seen by each profile (to have even more targeted recommendations) ranges between 70 and 130.

Although Profiles 3 and 6 are freshly created and will have no history when we begin collecting suggested abortion videos, they are still distinct from using the API or a private browser. This is because they will remain connected to YouTube while watching abortion-related content, allowing YouTube to provide us with a higher degree of personalization. Additionally, it is important to note that they do not serve the same purpose, since the first three profiles are used to examine viewpoints in YouTube recommendations, whereas the final three profiles are used to assess the spread of misinformation.

\subsubsection{Search queries}

\label{sec_queries}

Each sequence of collected video starts with a search query. For the first three profiles, where our goal is to see the prevalence of pro-, neutral-, and anti-abortion videos in YouTube recommendations, we must be aware of not including any search query bias. For that purpose, we chose to have these profiles solely search for the term ``abortion'', because this is the most neutral word that we can use to look for videos that address the topic. However, for the last three profiles, we want to start our watch session from the results of biased search queries, since YouTube claims to have eliminated abortion misinformation from the platform (as discussed in \ref{motivation}). For that purpose, we run search queries on the following sentences: ``abortion affects fertility'', ``abortion causes emotional and  mental harm'', ``abortion causes cancer'', ``abortion herbs are safer'', ``post-abortion syndrome'', ``died from abortion'', ``abortion is dangerous'', ``abortion destroys'', ``how to abort fetus'', ``illegal abortion'', ``abortion pills'', ``abortion is not safe'', ``self abortion'', ``home abortion'', ``abortion causes'', ``does mugwort induce abortion'', ``abortion reversal pills''.

Following the execution of the selected search queries, each profile scrolls the result page enough times to compile a list of videos that will serve as seed videos. Each profile then start its watching session from its seed videos, prompting YouTube to recommend videos. Since we are only interested in bias induced by the RS, we do not watch the seed videos in the order they appear on the search result page, to avoid any ranking bias from the search engine.

\subsubsection{Video selection}

We want to simulate individuals who are exclusively interested in abortion videos. A word frequency study revealed that the word ``abortion'' predominates in the titles (Over 80\% of videos that appear in the search results, after executing the search queries presented in Section~\ref{sec_queries}, include the word ``abortion'' in their titles.), and even when other common words like ``ban'' and ``rights'' are included in the title, the word ``abortion'' is also used in that title. For this reason, we only consider videos whose titles mention the word ``abortion''.

Using the same profile, each video is watched for an average of 10 seconds, and never watched twice to avoid having a different list of recommendations for the same video. Meanwhile, we collect video information like the number of views, duration, and so on. Most importantly, we scroll the page enough times to collect (in the order they appear in the list) recommended videos that include ``'abortion'' in the title. These videos are watched in the next iteration, and the procedure is repeated.

\subsection{Data annotation and analysis}

In similar empirical studies, the annotation of collected data represents a crucial part. But given the amount of data, it requires a lot of time and labor. Some studies employ machine learning (ML) \cite{misinfo1, homepage}, although such approaches necessitate human labeling of a portion of the data to be used as a training set, since there is hardly any dataset resembling the collected data. Moreover, approaches relying on ML may be inaccurate without a sufficiently large initial training set. As a result, we propose a method that only requires manual labeling, but meaningfully extracts most relevant items to be labelled.

First, we generate a recommendation graph where nodes represent videos, and a directed edge from a node A to a node B is added if video B appears in the recommendation list of video A. The weight of the edge from node A to node B reflects the rank of video B in A's list, and what we assume is the probability of watching video B next. Thus, we want the sum of a node's outgoing edges to equal one. Additionally, we do not want to have a big difference between the outgoing edges weights of a node, because we assume that a user will still be interested in a video that appears at the bottom of the list. In our case, when watching a video, an average of 8 videos with the word ``abortion'' in the title get recommended. Thus, we chose to assign weights according to a geometric series: $ \frac{1-r}{1-r^n} \times r^j $, such that $r=0.9$, $j$ is the rank of the video in the recommendation list, and $n$ is the length of the list. As a result, the weights for the videos at the bottom of the list will not be negligible. 

Then, we remove videos that were not recommended (i.e., seed videos that did not appear in any watched video suggestion list) and then calculate multiple measures for every node (in-degree, eccentricity\dots). We use six of them to rank nodes according to their importance in the graph:
\begin{itemize}
    \item \emph{In-degree}, which qualifies how many times a video was recommended.
    \item \emph{Weighted in-degree}, to take into account the rank of the video in others recommendation lists, because appearing few times at the top of the lists may be better than appearing many times at the bottom of the list.
    \item \emph{Eigen centrality}, which is related to the frequency with which a node is visited during a random walk.
    \item \emph{PageRank}, which uses the in-degree as the main measure, without letting nodes with no outgoing edges absorb all the scores of nodes connected to them.
    \item \emph{Katz centrality}, to enable popular videos to propagate their full weight to successors.
    \item \emph{Authority} (Hyperlink Induced Topic Search \cite{hits}), which gives a node a high score if it is linked by a node that links to many relevant vertices.
\end{itemize}

Since we want all measures to have the same weight in determining the significance of the videos, we normalize each measure and sum them to have a new score for each node. Furthermore, the measures are correlated in our graphs, and thus, averaging them makes sense: if they were not, the top videos (according to the averaged score) may not be considered important by any measure.

We rank videos according to the newly calculated score, and choose the top one percent from each of the six graphs to be labeled manually. We consider that this is more efficient than labeling all videos, not only because it is way quicker, but also because any genuine user who follows the recommendations will not watch thousands of videos, but only a very tiny subset, and their recommendation feed will probably include many of the most influential nodes. In other words, quantifying the bias of the RS using all suggested videos may not be the best solution, because recommending a biased item multiple times (in particular, in the recommendation list of popular items) may be way more dangerous than having many rarely recommended biased items.

Selected videos from the first three profiles were categorized as ``pro-abortion'', ``neutral'', or ``anti-abortion''. For the last three profiles, we classified them into: ``debunk'' ``misinformation'', ``neutral'', or ``misinformation''.

\subsection{Labeling methodology}

The labeling of abortion-related videos is carried out by first analyzing the text of the video's title, description, and popular well-rated comments to get an initial impression of the video's stance on the abortion topic. Determining these stances involves different methods, such as analyzing the titles, descriptions, and tags of their videos, reviewing the content of their videos for expressed opinions on abortion, examining comments and engagement feedback, researching their online presence and reputation, and checking for affiliations or endorsements with organizations holding known stances on abortion. Combining these approaches allows us to gain insights into the pro or anti-abortion tendencies of the responsible party behind the YouTube channel. In this way, each video is carefully watched and investigated, to determine whether it presents a clear pro-abortion or anti-abortion stance. If the video information strongly advocate for one side or the other, it is labeled accordingly. Alternatively, if no argument has been made for or against abortion, the video can be classified as neutral. 

In case of confusing content (difficult-to-classify videos), we further consider the source that produced the video's content to gain further insights into the potential biases and perspectives presented in the video. In this case, we mainly investigate the person or entity (i.e., the YouTube channel)  that uploaded the video, to see if they are known for promoting a particular type of pro or anti-abortion stance. We also analyze their content history, which may suggest a particular ideological leaning to abortion-related topics. Additionally, we also considered the time spent discussing each viewpoint, to strengthen our belief about the right classification of the video.

\section{Results \& Discussion}

\begin{figure}[ht]
  \centering
  \subfigure[Profile 1]{\includegraphics[width=0.32\textwidth]{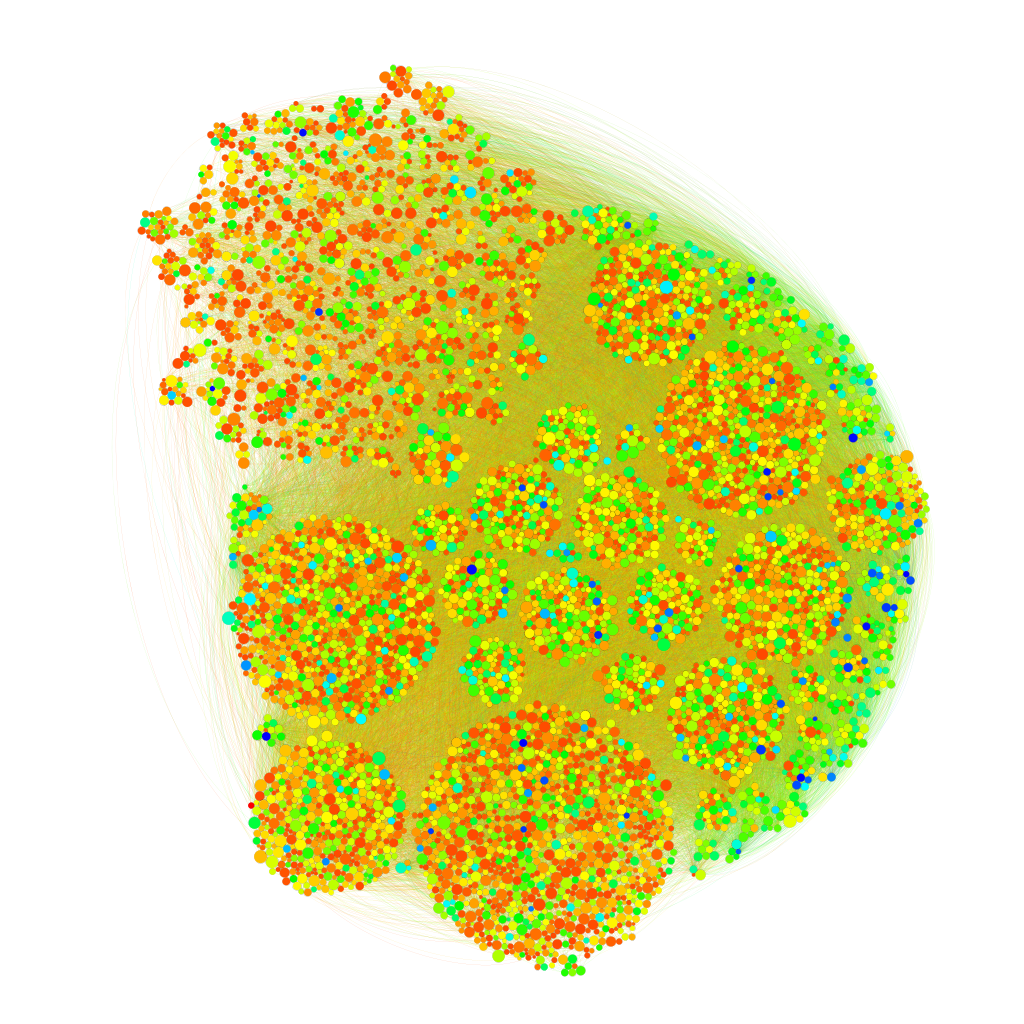}} 
  \subfigure[Profile 2]{\includegraphics[width=0.32\textwidth]{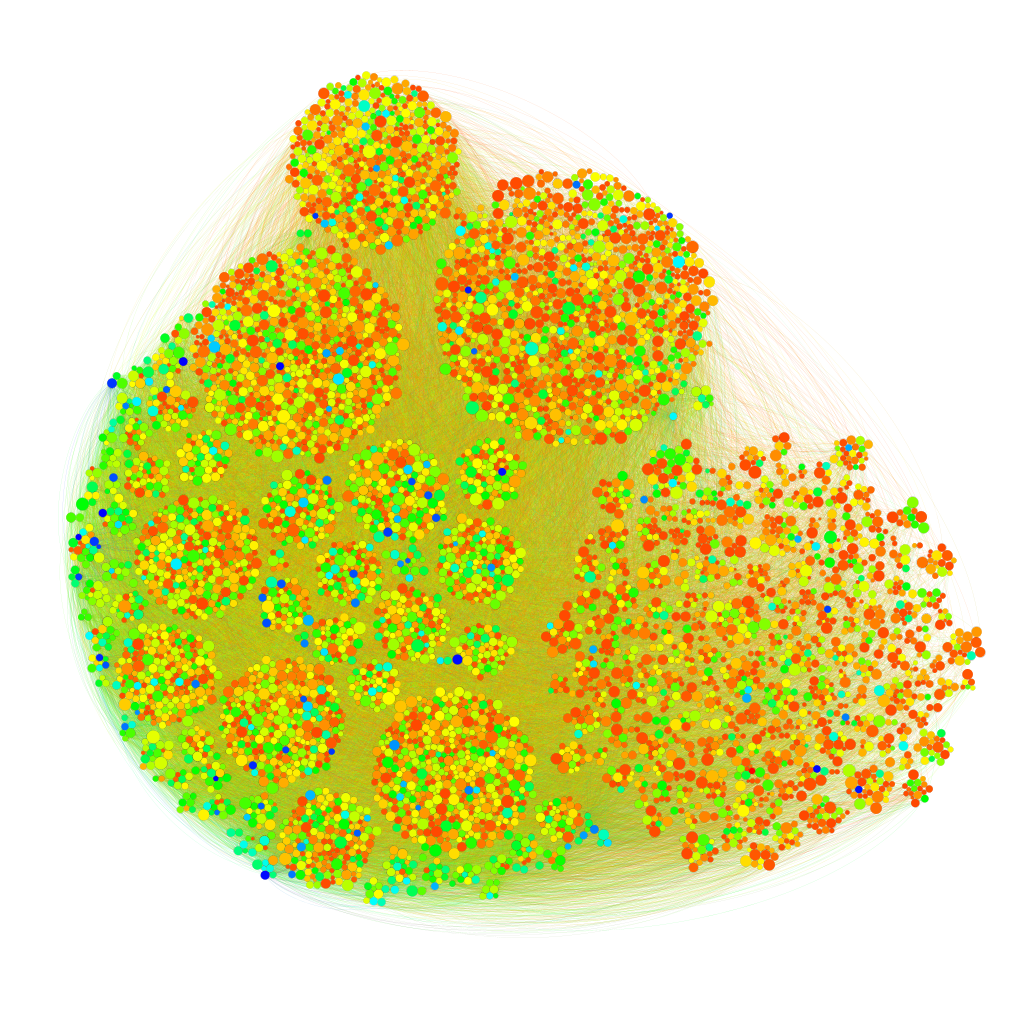}} 
  \subfigure[Profile 3]{\includegraphics[width=0.32\textwidth]{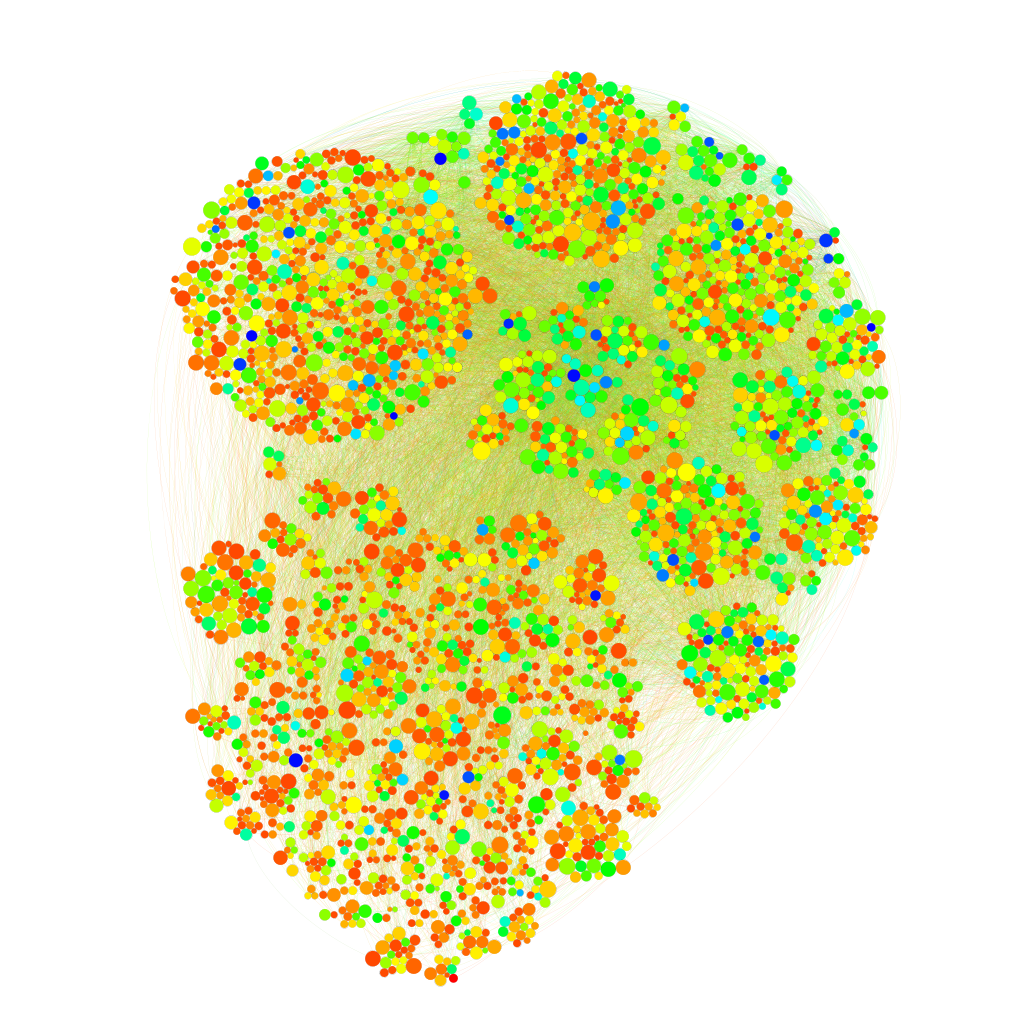}}
  \subfigure[Profile 4]{\includegraphics[width=0.32\textwidth]{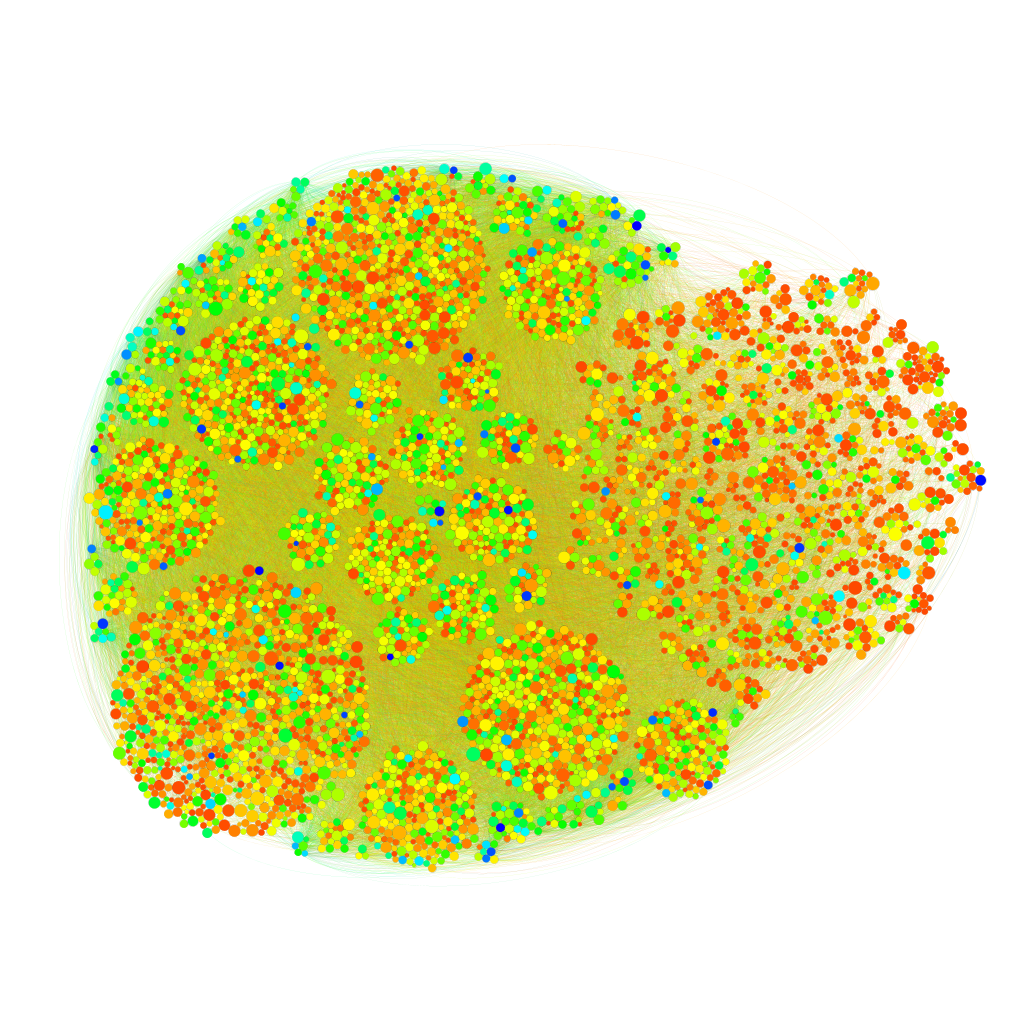}}
  \subfigure[Profile 5]{\includegraphics[width=0.32\textwidth]{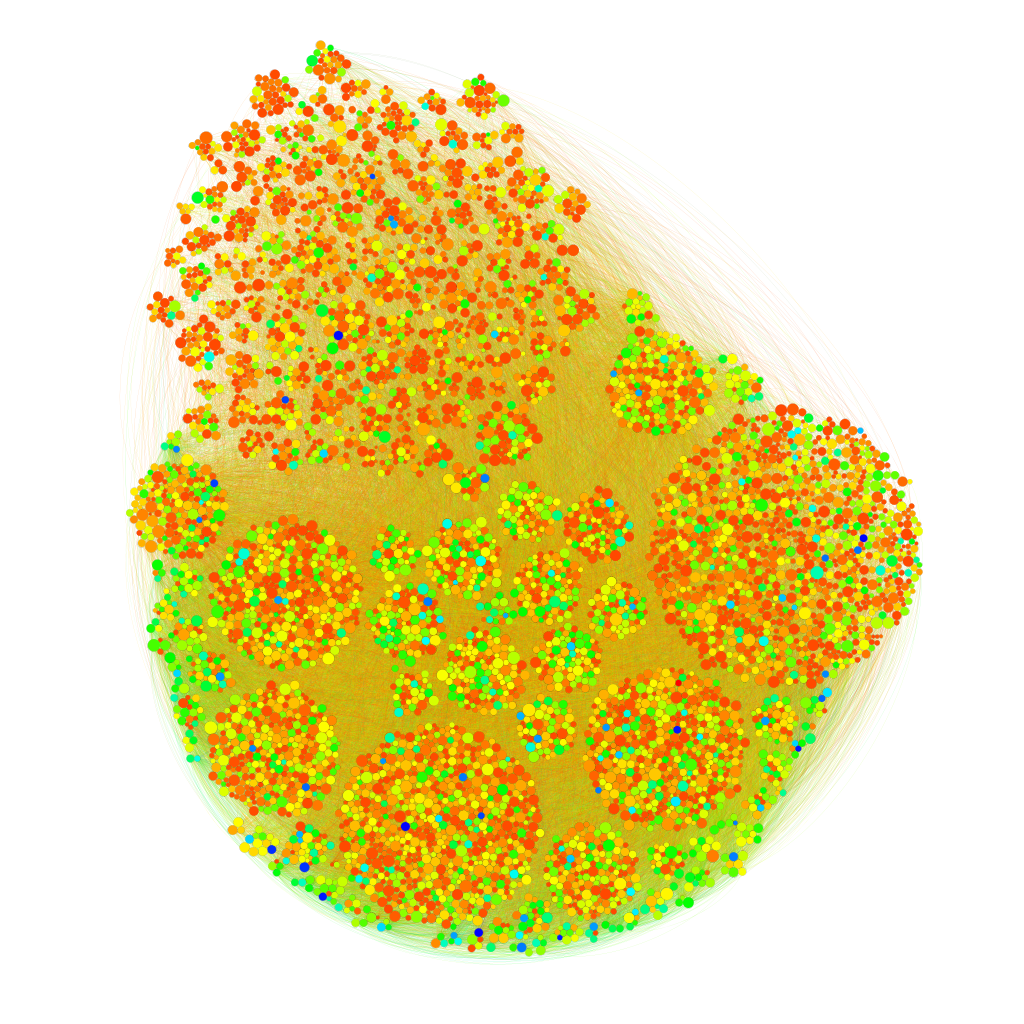}}
  \subfigure[Profile 6]{\includegraphics[width=0.32\textwidth]{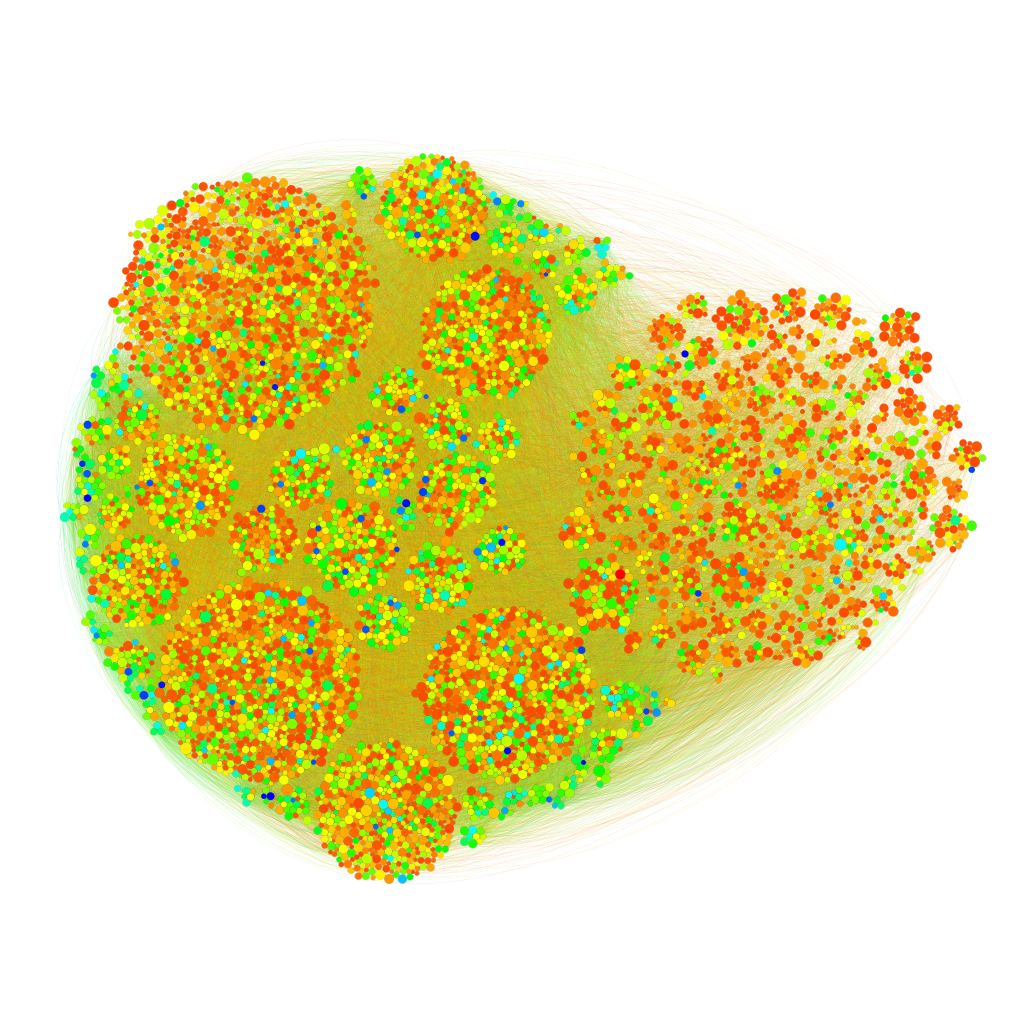}}
  \subfigure{\includegraphics[width=0.45\textwidth]{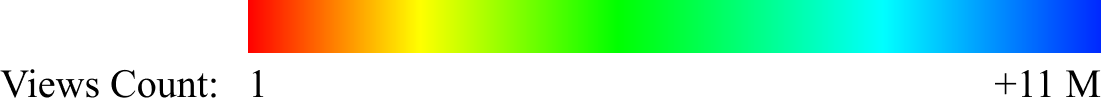}}
  \caption{Recommendations graphs for the six profiles. The node's size and color are correlated with the length and number of views of the video, respectively.}
\label{Graphs}
\end{figure}

\subsection{Classification of YouTube's recommendations}

A circle-pack layout was used to visualize the networks (Figure \ref{Graphs}). This arrangement enables nodes (i.e., videos) to be grouped hierarchically by attributes and plotted in circles, wherein nodes have close attribute values. In our case, nodes were grouped first by their ``in'' degree, and then by their ``weighted in'' degree.  The view count and the length of the video had no correlation with the six measures we used. A visual inspection of the graphs confirms this, since each circle has nodes of various sizes and colors. Table \ref{basicstats} reports some basic statistics about the networks. Except for the third profile, all graphs are comparable w.r.t. these statistics.

\begin{table}[H]
    \centering
    \setlength{\tabcolsep}{1em}
    \begin{tabular}{c c c c c c}
    \hline
    Recommendation graph & Nodes & Edges & Avg. degree & Avg. path length & Diameter \\
    \hline
    Profile 1 & 6837 & 53187 & 7.78 & 5.95 & 19 \\
    Profile 2 & 6974 & 55432 & 7.95 & 5.95 & 18 \\
    Profile 3 & 3241 & 14314 & 4.42 & 6.99 & 26 \\
    Profile 4 & 5761 & 43171 & 7.49 & 6.34 & 18 \\
    Profile 5 & 6734 & 47650 & 7.08 & 6.40 & 22 \\
    Profile 6 & 7486 & 54474 & 7.28 & 6.38 & 20 \\
    \hline
    \end{tabular}
\caption{Statistics of the recommendation graphs of the six profiles.}
\label{basicstats}
\end{table}

\begin{figure}[h]
  \centering
  \subfigure{\includegraphics[width=0.2\textwidth]{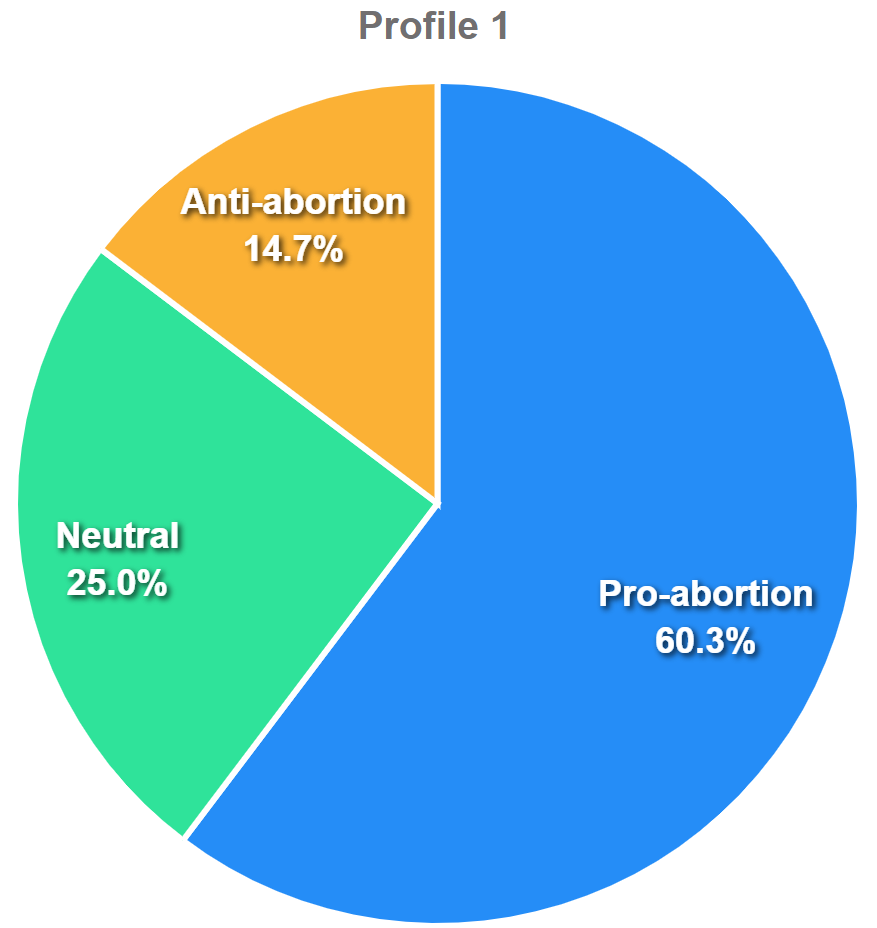}} 
  \hspace{10mm}
  \subfigure{\includegraphics[width=0.2\textwidth]{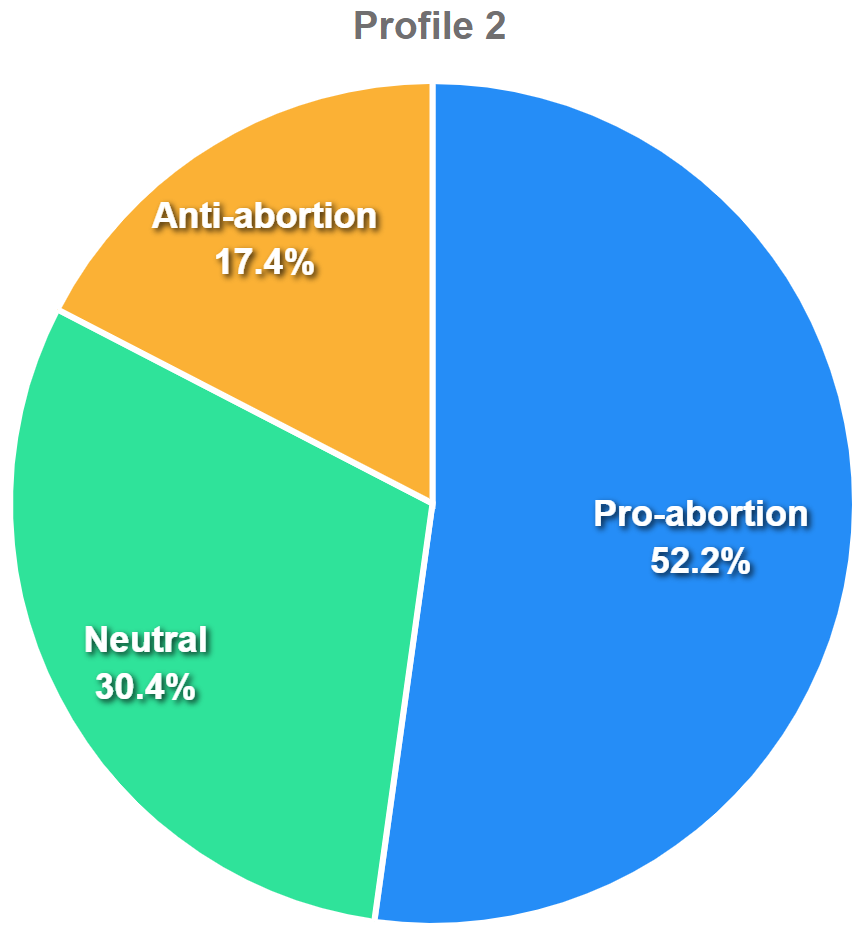}}
  \hspace{10mm}
  \subfigure{\includegraphics[width=0.2\textwidth]{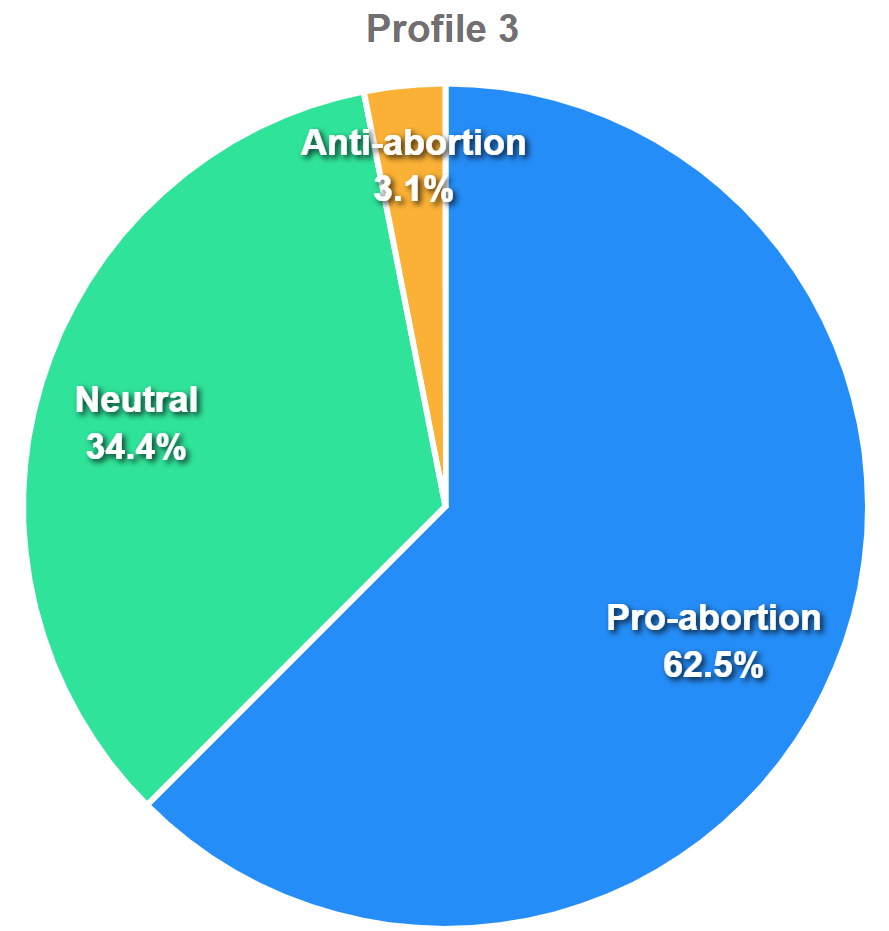}}
  \hspace{10mm}
  \subfigure{\includegraphics[width=0.2\textwidth]{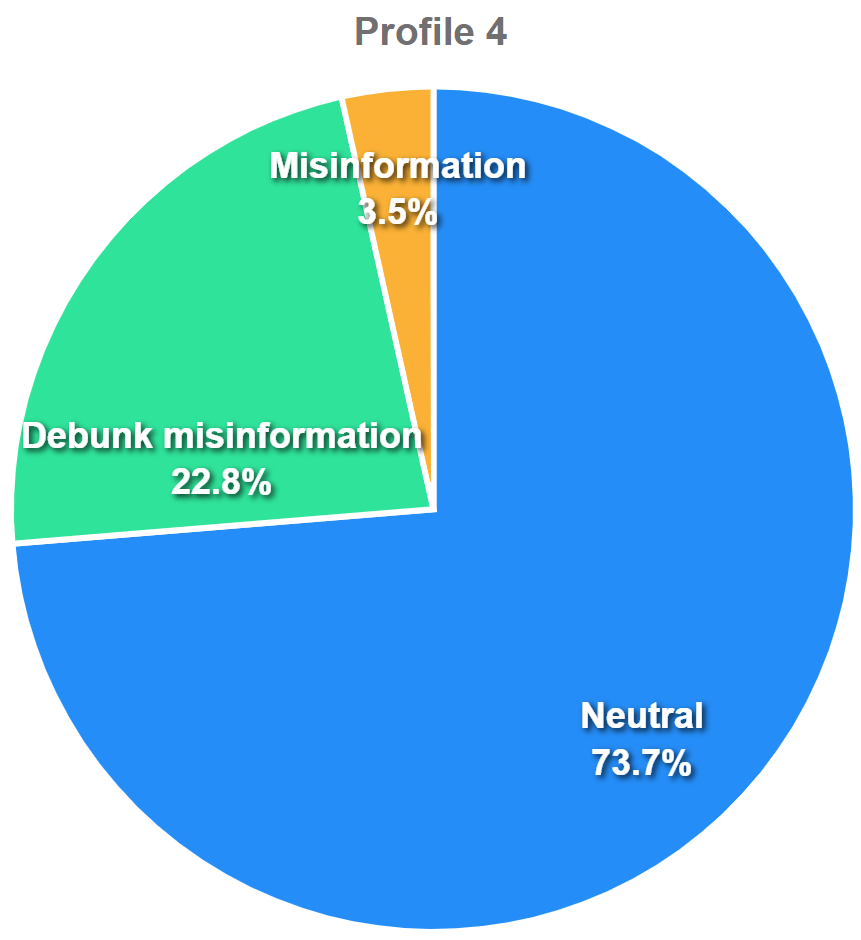}}
  \hspace{10mm}
  \subfigure{\includegraphics[width=0.2\textwidth]{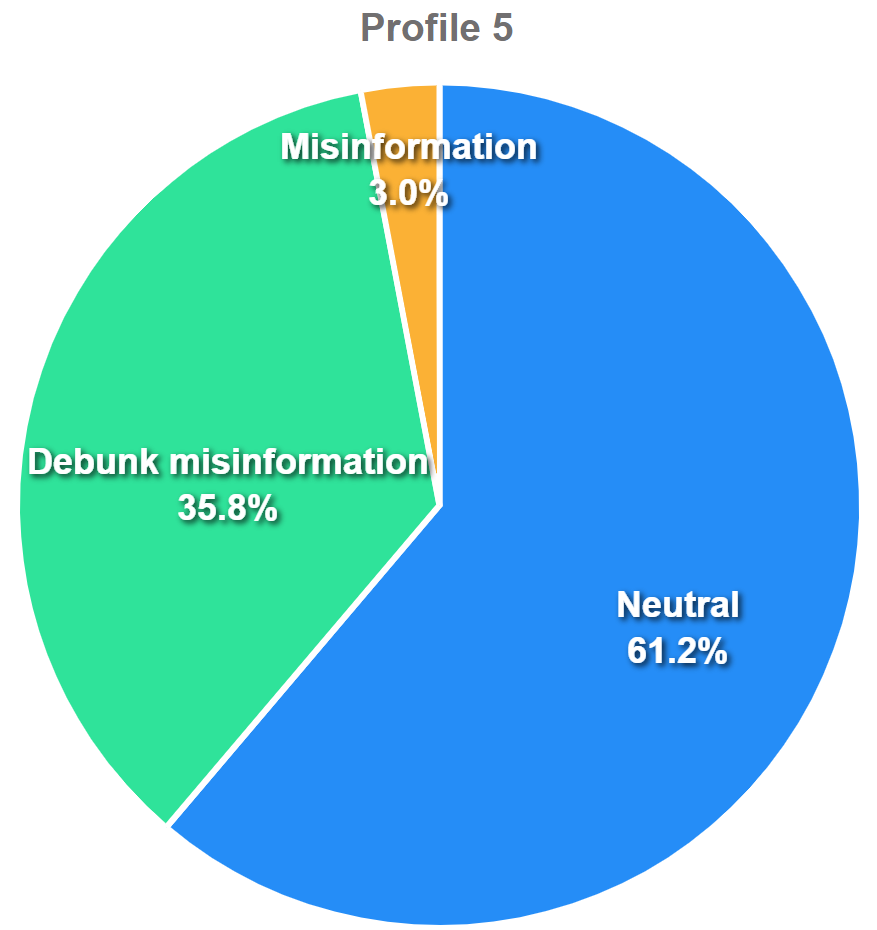}}
  \hspace{10mm}
  \subfigure{\includegraphics[width=0.2\textwidth]{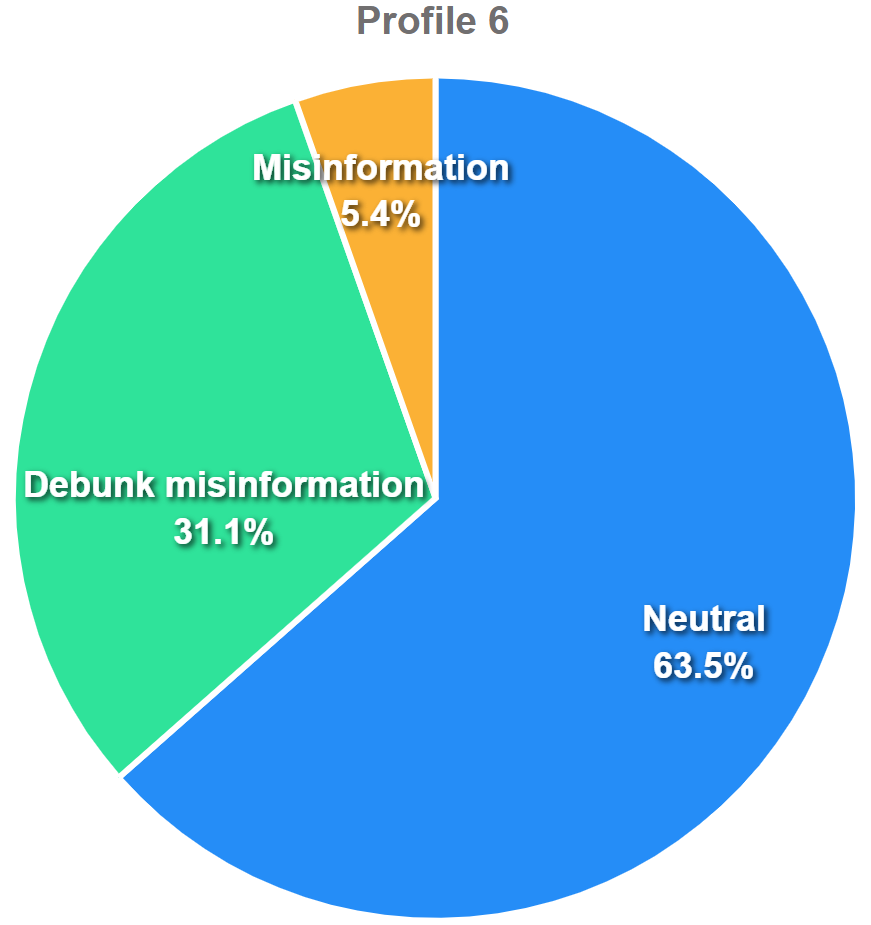}}
  \caption{Classification of the most influential videos in each graph.}
  \label{fig:piechart}
\end{figure}
  

After annotating the most influential videos in each network, we report in Figure \ref{fig:piechart} the percentage of each class. We found that the videos that endorse abortion are predominant in the 1\% of the videos most promoted by YouTube recommendations as identified by our ranking and selection procedure.

Compared to Profile 2, that was more exposed to anti-feminist content, Profile 1 was suggested a slightly greater proportion of pro-abortion videos, and a smaller proportion of anti-abortion videos. Profile 3, that has no viewing history, appears to rarely get anti-abortion suggestions.

In terms of dissemination of misinformation, YouTube seems to be successful in stopping the spread of false content, since all three profiles hardly ever receive suggestions for misleading content. Although the predominant recommendations are neutral, a huge chunk of the most influential videos are those that debunk misinformation. However, compared to the other two profiles, Profile 4 appears to receive the fewest recommendations for videos that dispel misconceptions. 

\subsection{Evaluation of the bias of YouTube's recommendation system}

We want to know how the top videos are ranked, and to have a single measurement of how biased each profile's recommendations are. For this purpose, we assign each labeled video a bias score $s$, which is $-1$ if it is ``pro-abortion'' or ``debunks misinformation'', $0$ if it is neutral, and $+1$ if it is ``anti-abortion'' or ``deceptive''. We then calculate the total bias for each profile, using the formula: $ \sum_{j=0}^{n-1} \frac{2 \times (n-j)}{n(n+1)} \times s_j$, where $n$ is the number of top 1\% videos, and $s_j$ is the bias score given to the video of rank $j+1$. The closer the measure is to $1$, the greater the influence of anti-abortion/deceptive videos in the profile; conversely, the closer it is to $-1$, the more ``pro-abortion''/``debunks misinformation'' videos are in the top.

\begin{table}[H]
    \centering
    \setlength{\tabcolsep}{1em}
    \begin{tabular}{c  c}
        \hline 
        Profile & Total bias \\
        \hline
        Profile 1 & -0.35 \\
        Profile 2 & -0.42 \\
        Profile 3 & -0.63 \\
        Profile 4 & -0.22 \\
        Profile 5 & -0.32 \\
        Profile 6 & -0.29 \\
        \hline
    \end{tabular}
\caption{Measure of existing bias in each profile.}
\label{bias}
\end{table}
 
The calculated measure, Table \ref{bias}, confirms that YouTube is more likely to recommend pro-abortion videos and those that debunk misinformation, since it yields a negative score for all profiles. However, when considering how videos are ranked, Profile 2 seems to tilt more towards pro-abortion videos than Profile 1. Indeed, while Profile 1 was offered more pro-abortion videos and fewer anti-abortion videos than Profile 2, the anti-abortion videos were more influential in its network.

To further compare the profiles (and not only the most important videos), we calculate the overlap coefficient (OC), which is the percentage of videos that are shared by the profiles (Table \ref{overlap}). However, the OC does not take into account how these common videos are ranked in the two profiles. As a result, we compute the Rank-biased Overlap \cite{rbo} (RBO) as well (Table \ref{overlap}). RBO analyzes two ranked lists, and produces a numeric value between 0 and 1 to indicate how similar they are. RBO is tuned for top-weightedness (i.e.,  similarities at the top of the two lists have a greater weight), which is adjusted by a parameter $p$ that controls the steepness of the weight decline. The smaller $p$, the more top-weighted the measure is. We used $p=0.97$ to limit the measure to only the first 1000 rankings when comparing profiles, and to give 95\% of the evaluation weight to the top 1\% of videos.

\begin{table}[H]
    \centering
    \setlength{\tabcolsep}{1em}
    \begin{tabular}{c c c}
    \hline
        & OC & RBO \\
        \hline
        Profiles 1 \& 2 & 0.78 & 0.47 \\
        Profiles 1 \& 3 & 0.83 & 0.23 \\
        Profiles 2 \& 3 & 0.84 & 0.40 \\
        Profiles 4 \& 5 & 0.77 & 0.44 \\
        Profiles 4 \& 6 & 0.81 & 0.50 \\
        Profiles 5 \& 6 & 0.80 & 0.64 \\
    \hline
    \end{tabular}
\caption{Overlap between profiles.}
\label{overlap}
\end{table}

Profiles that started with contrasted watch histories (Profiles 1 and 2 / Profiles 4 and 5) had slightly fewer common videos when compared with those that started with no watch history. However, we can say that all profiles had a lot of similar recommendations. When considering the influence of the overlapping videos in their networks, the RBO measure shows that, despite the considerable overlap, videos are not recommended in the same way, particularly in the top rankings.

We found no substantial differences across profiles in the proportion of each class for labeled videos, except for the lack of anti-abortion recommendations for Profile 3. This might be because the RS treats all profiles similarly when proposing abortion videos, or because the quantity of videos seen to establish the profile history was insignificant in comparison to the number of watched videos that discuss abortion. In other words, even if the watch history may have had a bigger effect on the first recommended videos, this effect may have diminished as the experiment went on. That is why, while studying the effects of browsing/watch history on the RS, we recommend to periodically revisit a collection of chosen items, to ensure that the effect of browsing/watch history persists even in the latter stages of the experiment.

\section{Discussion}
\subsection{Review of findings and contributions}

In this work, we focused on examining YouTube recommendations with the aim of auditing the spread of misinformation and biased content pertaining to abortion, a complex and multifaceted topic encompassing both social and health dimensions. We acquired 11174 unique videos which appeared over 38k times in the recommendations given by YouTube to six simulated individuals from various backgrounds looking for abortion content on the platform. We built a graph of recommendations for each profile, and through the use of graph theory, we annotated the most important videos in the network based on averaging different centrality metrics. Our choice of centrality metrics (In-degree, Weighted in-degree, Eigen centrality, PageRank, Katz centrality, Authority) demonstrates a thoughtful consideration of different dimensions of influence, such as connectivity, authority, and popularity. The average of these metrics allows to (1) identify videos that have consistently high influence across various criteria, enhancing the reliability of our ranking, and (2) mitigate potential biases that might arise from relying on a single measure, and provide a more comprehensive view of the videos' importance within the network. 

The idea of using graph centrality measures to select the most important nodes (videos) is a highly commendable approach to performing bias analysis, which is often used to evaluate social influence in social networks \cite{ISHFAQ20229376,8692729}. On the one hand, by identifying these top-ranked videos, we are effectively identifying content that significantly shapes the network's dynamics and user interactions. These influential videos are likely to hold substantial effect over opinions and information dissemination, perhaps perpetuating or challenging biases. On the other hand, this would emphasize a strategic and optimized use of resources and efforts. Thus, our approach circumvents the labeling bottleneck, which is usually challenging and time-consuming, especially for large video datasets. This optimization results in a notably faster auditing process, enhancing the overall efficiency. We underline the importance of training the sock puppets, including the initial training to tailor the selected themes and ideologies, because trained sock puppets offer a balance between fake and real users for audit purposes \cite{searchbias,10.1007/978-3-030-76228-5_8}. More precisely, they trigger the same feedback loops as real users to allow for better personalization, while isolating the impact of YouTube's recommendations on bias and radicalization. This is difficult to do if we only use real user interaction histories.
 
We anticipated considerable variations in the recommendations that users with contrasted watching histories would receive, and yet, YouTube appears to provide similar recommendations to all users when they search for and watch abortion-related videos. We can thus claim that (1) pro-abortion content seems to be more prominent than anti-abortion content in YouTube's recommendations, and (2) YouTube has been successful in halting the dissemination of false information, at least when it comes to recommendations. In this context, our research findings pertaining to the dissemination of abortion-related misinformation on YouTube align with the platform's announcement in July 2022 that it intends to undertake measures to eliminate such misinformation \cite{tweet,cnn}. Notably, our study was conducted subsequently to this announcement. It is worth mentioning that, for a more robust validation of YouTube's announcement, a parallel study conducted prior to the aforementioned date could have provided a basis for enhanced comparison and confirmation.
Furthermore, our analysis show that users with previous watching (and searching) histories of medical and medicine-related content tend to encounter fewer videos dispelling abortion misconceptions, while individuals who previously watched feminism-related content appear to receive more anti-abortion videos. With all that being said, our contribution stands as the first study to present these auditing investigations and results  regarding YouTube abortion-related recommendations.

\subsection{Limitations and future directions}

The scope of the study was restricted to measuring bias in the recommendation list using the topmost 1\% of influential videos selected by our graph analysis approach. While these findings provide valuable insights, particularly in the context of our abortion-focused study, it is important to note that the applicability of these results' production details may not be universally applicable to other contentious topics in an identical manner. The feasibility and effectiveness of employing similar methodologies for different topics will inevitably depend on various factors, including the topic's sensitivity, its impact on individuals and societies, and the nature, volume, and prevalence of its related content across the internet. These factors will definitely influence the ranking of videos  (or information sources in general) based on the centrality measurements, which will give further insights on the considerations and selection procedures of the most influential videos. Thus, we do not consider this variability as a critical limitation of our study, because it basically serves as a parameter that needs to be tailored to each specific use case when implementing such auditing methodologies. However, we agree that using 1\% of the videos for analysis could potentially limit the scope and generalizability of the findings. In this context, it would be interesting for future works to employ some machine learning techniques to extrapolate from partially labeled data, which can help to address this limitation by expanding the dataset. Thus, comparing the outcomes of analyses performed on both partially labeled data and ML-extrapolated data could provide useful insights regarding the impact of data size and the performance of the extrapolation process, which could enhance the comprehensiveness and reliability of the conclusions of our study.

In addition, the data labeling process was conducted by an individual committed to maintaining strict neutrality, objectivity, and accuracy with repetitive revision. Despite these efforts, the potential for some bias and errors cannot be entirely ruled out. Even with the best intentions, individual interpretations and subtle subjectivity can inadvertently seep into the labeling process. Therefore, we agree that our study's robustness could be further enhanced by increasing the number of labeled videos and adopting a more rigorous labeling approach (e.g., using a collaborative labeling method or implementing comprehensive quality control checks). These improvements would lead to more precise measurements and statistically strong results.

Our methodology can be used by any third-party auditor, regardless of platform or topic, and it can be swiftly adopted to audit any RS, without any understanding of the underlying algorithm. However, to investigate the formation of filter bubbles \cite{filterbubble} in recommendations (or any type of radicalization that may happen over time), the annotation of the entire collected data may be required, since we cannot examine such dynamic properties with only a few labeled items. An interesting  extension to this work might also include measuring bias in the homepage feed and the search engine.

\section{Conclusion}

This study sheds light on an important problem concerning the algorithmic personalization behind YouTube's recommendations for abortion-related videos, and the impact of inherent recommendation biases on the dissemination of misinformation and biased content. The importance of this research is heightened by the increased reliance on internet information, which is driven by its accessibility and cost-effectiveness. This tendency is especially noticeable in countries with low and middle incomes, where social media are increasingly used for health-related purposes \cite{HAGG201892}. As a result, the spread of health misinformation on platforms such as YouTube has the potential to lead to harmful abortion-related health decisions that can have serious consequences for people's lives.

The study's findings show that YouTube's recommendations tend to prioritize pro-abortion content over anti-abortion content, and the platform has generally demonstrated efficacy in reducing misinformation spread within its recommendations. However, some findings of this study raise intriguing possibilities regarding the potential emergence of polarization and filter bubble phenomena. Indeed, the observation that users who engage with medical content are exposed to fewer videos dispelling abortion misconceptions, while those interested in feminism-related content receive more anti-abortion videos, could inadvertently contribute to these concerns. Such behavior may reinforce pre-existing beliefs, contributing to polarization and inhibiting constructive discourse. Additionally, the identified content distribution patterns could imply the presence of filter bubbles, which limit users' exposure to a diverse range of perspectives. Recommendation systems have to judiciously personalize their recommendations according to user preferences while respecting diversity and fairness criteria. These safety measurements in the recommendations are crucial for preventing the emergence of polarization and filter bubble effects \cite{10.1145/2566486.2568012,10.1007/978-3-031-21743-2_53}.
    
In conclusion, by conducting this analysis, we sought to provide insights on the potential impact of algorithmic recommendations in shaping public perceptions and understanding of abortion-related content on social media platforms, which heavily rely on recommendation algorithms. Our research aimed to identify and address any inaccuracies or biases that could contribute to the spread of misinformation, with the goal of promoting a more reliable and trustworthy information environment regarding abortion and its related healthcare implications.

\subsection*{Statements and Declarations}
\paragraph{Competing interests} The authors declare that they have no known competing financial interests or personal relationships that could have appeared to influence the work reported in this paper.
\paragraph{Funding} The authors did not receive support from any organization for the submitted work.
\paragraph{Ethical approval} This research involving the collection and analysis of data obtained from the YouTube API was conducted in compliance with all relevant ethical guidelines and regulations of YouTube API Services Policies \url{https://developers.google.com/youtube/terms/developer-policies}.

\bibliographystyle{plainurl}
\bibliography{sample}

\begin{thebibliography}{10}

\bibitem{YTsearch}
Find videos faster - youtube help.
\newblock (last accessed December 19, 2022).
\newblock URL: \url{https://support.google.com/youtube/answer/9872296}.

\bibitem{YTguidlines}
Youtube community guidelines \& policies - how youtube works.
\newblock (last accessed December 19, 2022).
\newblock URL: \url{https://www.youtube.com/howyoutubeworks/policies/community-guidelines/#community-guidelines}.

\bibitem{provac}
Deena Abul-Fottouh, Melodie~Yunju Song, and Anatoliy Gruzd.
\newblock Examining algorithmic biases in youtube’s recommendations of vaccine videos.
\newblock {\em International Journal of Medical Informatics}, 140:104175, 2020.
\newblock URL: \url{https://www.sciencedirect.com/science/article/pii/S1386505619308743}, \href {https://doi.org/10.1016/j.ijmedinf.2020.104175} {\path{doi:10.1016/j.ijmedinf.2020.104175}}.

\bibitem{abul2020examining}
Deena Abul-Fottouh, Melodie~Yunju Song, and Anatoliy Gruzd.
\newblock Examining algorithmic biases in youtube’s recommendations of vaccine videos.
\newblock {\em International Journal of Medical Informatics}, 140:104175, 2020.

\bibitem{YTstats}
Salman Aslam.
\newblock youtube by the numbers (2022): Stats, demographic \& fun facts, Aug 2022.
\newblock (last accessed 19 December, 2022).
\newblock URL: \url{https://www.omnicoreagency.com/youtube-statistics/}.

\bibitem{10.1145/3209581}
Ricardo Baeza-Yates.
\newblock Bias on the web.
\newblock {\em Commun. ACM}, 61(6):54–61, may 2018.
\newblock \href {https://doi.org/10.1145/3209581} {\path{doi:10.1145/3209581}}.

\bibitem{doi:10.1126/science.aaa1160}
Eytan Bakshy, Solomon Messing, and Lada~A. Adamic.
\newblock Exposure to ideologically diverse news and opinion on facebook.
\newblock {\em Science}, 348(6239):1130--1132, 2015.
\newblock URL: \url{https://www.science.org/doi/abs/10.1126/science.aaa1160}, \href {https://arxiv.org/abs/https://www.science.org/doi/pdf/10.1126/science.aaa1160} {\path{arXiv:https://www.science.org/doi/pdf/10.1126/science.aaa1160}}, \href {https://doi.org/10.1126/science.aaa1160} {\path{doi:10.1126/science.aaa1160}}.

\bibitem{algoaudit}
Jack Bandy.
\newblock Problematic machine behavior: A systematic literature review of algorithm audits.
\newblock {\em Proc. ACM Hum.-Comput. Interact.}, 5(CSCW1), apr 2021.
\newblock \href {https://doi.org/10.1145/3449148} {\path{doi:10.1145/3449148}}.

\bibitem{bartley2021auditing}
Nathan Bartley, Andres Abeliuk, Emilio Ferrara, and Kristina Lerman.
\newblock Auditing algorithmic bias on twitter.
\newblock In {\em Proceedings of the 13th ACM Web Science Conference 2021}, pages 65--73, 2021.

\bibitem{8692729}
Ayan~Kumar Bhowmick, Martin Gueuning, Jean-Charles Delvenne, Renaud Lambiotte, and Bivas Mitra.
\newblock Temporal sequence of retweets help to detect influential nodes in social networks.
\newblock {\em IEEE Transactions on Computational Social Systems}, 6(3):441--455, 2019.
\newblock \href {https://doi.org/10.1109/TCSS.2019.2907553} {\path{doi:10.1109/TCSS.2019.2907553}}.

\bibitem{10.1145/2716281.2836098}
Juan~Miguel Carrascosa, Jakub Mikians, Ruben Cuevas, Vijay Erramilli, and Nikolaos Laoutaris.
\newblock I always feel like somebody's watching me: Measuring online behavioural advertising.
\newblock In {\em Proceedings of the 11th ACM Conference on Emerging Networking Experiments and Technologies}, CoNEXT '15, New York, NY, USA, 2015. Association for Computing Machinery.
\newblock \href {https://doi.org/10.1145/2716281.2836098} {\path{doi:10.1145/2716281.2836098}}.

\bibitem{10.1145/3219819.3220122}
Shi-Yong Chen, Yang Yu, Qing Da, Jun Tan, Hai-Kuan Huang, and Hai-Hong Tang.
\newblock Stabilizing reinforcement learning in dynamic environment with application to online recommendation.
\newblock In {\em Proceedings of the 24th ACM SIGKDD International Conference on Knowledge Discovery \& Data Mining}, KDD '18, page 1187–1196, New York, NY, USA, 2018. Association for Computing Machinery.
\newblock \href {https://doi.org/10.1145/3219819.3220122} {\path{doi:10.1145/3219819.3220122}}.

\bibitem{chou2018addressing}
Wen-Ying~Sylvia Chou, April Oh, and William~MP Klein.
\newblock Addressing health-related misinformation on social media.
\newblock {\em Jama}, 320(23):2417--2418, 2018.

\bibitem{WhoAudits}
Sasha Costanza-Chock, Inioluwa~Deborah Raji, and Joy Buolamwini.
\newblock Who audits the auditors? recommendations from a field scan of the algorithmic auditing ecosystem.
\newblock In {\em 2022 ACM Conference on Fairness, Accountability, and Transparency}, FAccT '22, page 1571–1583, New York, NY, USA, 2022. Association for Computing Machinery.
\newblock \href {https://doi.org/10.1145/3531146.3533213} {\path{doi:10.1145/3531146.3533213}}.

\bibitem{YTRecSys}
Paul Covington, Jay Adams, and Emre Sargin.
\newblock Deep neural networks for youtube recommendations.
\newblock In {\em Proceedings of the 10th ACM Conference on Recommender Systems}, RecSys '16, page 191–198, New York, NY, USA, 2016. Association for Computing Machinery.
\newblock \href {https://doi.org/10.1145/2959100.2959190} {\path{doi:10.1145/2959100.2959190}}.

\bibitem{cnn}
Clare Duffy.
\newblock Youtube will start removing misinformation related to abortion, Jul 2022.
\newblock (last accessed December 19, 2022).
\newblock URL: \url{https://edition.cnn.com/2022/07/21/tech/youtube-abortion-misinformation-policy/index.html}.

\bibitem{10.1007/978-3-031-21743-2_53}
Zakaria El-Moutaouakkil, Mohamed Lechiakh, and Alexandre Maurer.
\newblock Polarization in personalized recommendations: Balancing safety and accuracy.
\newblock In Ngoc~Thanh Nguyen, Tien~Khoa Tran, Ualsher Tukayev, Tzung-Pei Hong, Bogdan Trawi{\'{n}}ski, and Edward Szczerbicki, editors, {\em Intelligent Information and Database Systems}, pages 661--674, Cham, 2022. Springer International Publishing.

\bibitem{10.1145/3372923.3404787}
Flavio Figueiredo, Felipe Giori, Guilherme Soares, Mariana Arantes, Jussara~M. Almeida, and Fabricio Benevenuto.
\newblock Understanding targeted video-ads in children's content.
\newblock In {\em Proceedings of the 31st ACM Conference on Hypertext and Social Media}, HT '20, page 151–160, New York, NY, USA, 2020. Association for Computing Machinery.
\newblock \href {https://doi.org/10.1145/3372923.3404787} {\path{doi:10.1145/3372923.3404787}}.

\bibitem{google}
Jen Fitzpatrick.
\newblock Protecting people's privacy on health topics, Jul 2022.
\newblock (last accessed December 19, 2022).
\newblock URL: \url{https://blog.google/technology/safety-security/protecting-peoples-privacy-on-health-topics/}.

\bibitem{10.1145/3437963.3441824}
Yingqiang Ge, Shuchang Liu, Ruoyuan Gao, Yikun Xian, Yunqi Li, Xiangyu Zhao, Changhua Pei, Fei Sun, Junfeng Ge, Wenwu Ou, and Yongfeng Zhang.
\newblock Towards long-term fairness in recommendation.
\newblock In {\em Proceedings of the 14th ACM International Conference on Web Search and Data Mining}, WSDM '21, page 445–453, New York, NY, USA, 2021. Association for Computing Machinery.
\newblock \href {https://doi.org/10.1145/3437963.3441824} {\path{doi:10.1145/3437963.3441824}}.

\bibitem{HAGG201892}
Emily Hagg, V.~Susan Dahinten, and Leanne~M. Currie.
\newblock The emerging use of social media for health-related purposes in low and middle-income countries: A scoping review.
\newblock {\em International Journal of Medical Informatics}, 115:92--105, 2018.
\newblock URL: \url{https://www.sciencedirect.com/science/article/pii/S1386505618304568}, \href {https://doi.org/10.1016/j.ijmedinf.2018.04.010} {\path{doi:10.1016/j.ijmedinf.2018.04.010}}.

\bibitem{HoangLouis}
L{\^e}~Nguy{\^e}n Hoang, Louis Faucon, and El-Mahdi El-Mhamdi.
\newblock Recommendation algorithms, a neglected opportunity for public health revue m\'edecine et philosophie.
\newblock 2021.

\bibitem{hosseinmardi2021examining}
Homa Hosseinmardi, Amir Ghasemian, Aaron Clauset, Markus Mobius, David~M Rothschild, and Duncan~J Watts.
\newblock Examining the consumption of radical content on youtube.
\newblock {\em Proceedings of the National Academy of Sciences}, 118(32):e2101967118, 2021.

\bibitem{searchbias}
Eslam Hussein, Prerna Juneja, and Tanushree Mitra.
\newblock Measuring misinformation in video search platforms: An audit study on youtube.
\newblock {\em Proc. ACM Hum.-Comput. Interact.}, 4(CSCW1), may 2020.
\newblock \href {https://doi.org/10.1145/3392854} {\path{doi:10.1145/3392854}}.

\bibitem{hussein2020measuring}
Eslam Hussein, Prerna Juneja, and Tanushree Mitra.
\newblock Measuring misinformation in video search platforms: An audit study on youtube.
\newblock {\em Proceedings of the ACM on Human-Computer Interaction}, 4(CSCW1):1--27, 2020.

\bibitem{ISHFAQ20229376}
Umar Ishfaq, Hikmat~Ullah Khan, and Saqib Iqbal.
\newblock Identifying the influential nodes in complex social networks using centrality-based approach.
\newblock {\em Journal of King Saud University - Computer and Information Sciences}, 34(10, Part B):9376--9392, 2022.
\newblock URL: \url{https://www.sciencedirect.com/science/article/pii/S1319157822003457}, \href {https://doi.org/10.1016/j.jksuci.2022.09.016} {\path{doi:10.1016/j.jksuci.2022.09.016}}.

\bibitem{beh1}
Baris Kirdemir and Nitin Agarwal.
\newblock Exploring bias and information bubbles in youtube's video recommendation networks.
\newblock In Rosa~Maria Benito, Chantal Cherifi, Hocine Cherifi, Esteban Moro, Luis~M. Rocha, and Marta Sales-Pardo, editors, {\em Complex Networks {\&} Their Applications X}, pages 166--177, Cham, 2022. Springer International Publishing.

\bibitem{api1}
Baris Kirdemir, Joseph Kready, Esther Mead, Muhammad~Nihal Hussain, and Nitin Agarwal.
\newblock Examining video recommendation bias on youtube.
\newblock In Ludovico Boratto, Stefano Faralli, Mirko Marras, and Giovanni Stilo, editors, {\em Advances in Bias and Fairness in Information Retrieval}, pages 106--116, Cham, 2021. Springer International Publishing.

\bibitem{hits}
Jon~M. Kleinberg.
\newblock Authoritative sources in a hyperlinked environment.
\newblock {\em J. ACM}, 46(5):604–632, sep 1999.
\newblock \href {https://doi.org/10.1145/324133.324140} {\path{doi:10.1145/324133.324140}}.

\bibitem{10.1145/3308558.3313682}
Huyen Le, Raven Maragh, Brian Ekdale, Andrew High, Timothy Havens, and Zubair Shafiq.
\newblock Measuring political personalization of google news search.
\newblock In {\em The World Wide Web Conference}, WWW '19, page 2957–2963, New York, NY, USA, 2019. Association for Computing Machinery.
\newblock \href {https://doi.org/10.1145/3308558.3313682} {\path{doi:10.1145/3308558.3313682}}.

\bibitem{ledwich2019algorithmic}
Mark Ledwich and Anna Zaitsev.
\newblock Algorithmic extremism: Examining youtube's rabbit hole of radicalization, 2019.
\newblock \href {https://arxiv.org/abs/1912.11211} {\path{arXiv:1912.11211}}.

\bibitem{api2}
Mark Ledwich and Anna Zaitsev.
\newblock Algorithmic extremism: Examining youtube’s rabbit hole of radicalization.
\newblock {\em First Monday}, 25(3), Feb. 2020.
\newblock URL: \url{https://firstmonday.org/ojs/index.php/fm/article/view/10419}, \href {https://doi.org/10.5210/fm.v25i3.10419} {\path{doi:10.5210/fm.v25i3.10419}}.

\bibitem{Lie008334}
Heidi Oi-Yee Li, Elena Pastukhova, Olivier Brandts-Longtin, Marcus~G Tan, and Mark~G Kirchhof.
\newblock Youtube as a source of misinformation on covid-19 vaccination: a systematic analysis.
\newblock {\em BMJ Global Health}, 7(3), 2022.
\newblock URL: \url{https://gh.bmj.com/content/7/3/e008334}, \href {https://arxiv.org/abs/https://gh.bmj.com/content/7/3/e008334.full.pdf} {\path{arXiv:https://gh.bmj.com/content/7/3/e008334.full.pdf}}, \href {https://doi.org/10.1136/bmjgh-2021-008334} {\path{doi:10.1136/bmjgh-2021-008334}}.

\bibitem{LITTMAN201419}
Lisa~L. Littman, Adam Jacobs, Rennie Negron, Tara Shochet, Marji Gold, and Miriam Cremer.
\newblock Beliefs about abortion risks in women returning to the clinic after their abortions: a pilot study.
\newblock {\em Contraception}, 90(1):19--22, 2014.
\newblock URL: \url{https://www.sciencedirect.com/science/article/pii/S0010782414001152}, \href {https://doi.org/10.1016/j.contraception.2014.03.005} {\path{doi:10.1016/j.contraception.2014.03.005}}.

\bibitem{10.1145/3397271.3401083}
Dugang Liu, Pengxiang Cheng, Zhenhua Dong, Xiuqiang He, Weike Pan, and Zhong Ming.
\newblock A general knowledge distillation framework for counterfactual recommendation via uniform data.
\newblock In {\em Proceedings of the 43rd International ACM SIGIR Conference on Research and Development in Information Retrieval}, SIGIR '20, page 831–840, New York, NY, USA, 2020. Association for Computing Machinery.
\newblock \href {https://doi.org/10.1145/3397271.3401083} {\path{doi:10.1145/3397271.3401083}}.

\bibitem{healthinfoonyoutube}
Kapil~Chalil Madathil, A~Joy Rivera-Rodriguez, Joel~S Greenstein, and Anand~K Gramopadhye.
\newblock Healthcare information on youtube: A systematic review.
\newblock {\em Health Informatics Journal}, 21(3):173--194, 2015.
\newblock PMID: 24670899.
\newblock \href {https://arxiv.org/abs/https://doi.org/10.1177/1460458213512220} {\path{arXiv:https://doi.org/10.1177/1460458213512220}}, \href {https://doi.org/10.1177/1460458213512220} {\path{doi:10.1177/1460458213512220}}.

\bibitem{muhammed2022disaster}
Sadiq Muhammed~T and Saji~K Mathew.
\newblock The disaster of misinformation: a review of research in social media.
\newblock {\em International journal of data science and analytics}, 13(4):271--285, 2022.

\bibitem{10.1145/3531146.3533136}
Preetam Nandy, Cyrus DiCiccio, Divya Venugopalan, Heloise Logan, Kinjal Basu, and Noureddine El~Karoui.
\newblock Achieving fairness via post-processing in web-scale recommender systems✱.
\newblock In {\em Proceedings of the 2022 ACM Conference on Fairness, Accountability, and Transparency}, FAccT '22, page 715–725, New York, NY, USA, 2022. Association for Computing Machinery.
\newblock \href {https://doi.org/10.1145/3531146.3533136} {\path{doi:10.1145/3531146.3533136}}.

\bibitem{newsReport}
Nic Newman, Richard Fletcher, Craig~T Robertson, Kirsten Eddy, and Rasmus~Kleis Nielsen.
\newblock Reuters institute digital news report 2022.
\newblock (last accessed December 19, 2022).
\newblock URL: \url{https://reutersinstitute.politics.ox.ac.uk/sites/default/files/2022-06/Digital_News-Report_2022.pdf}.

\bibitem{10.1145/2566486.2568012}
Tien~T. Nguyen, Pik-Mai Hui, F.~Maxwell Harper, Loren Terveen, and Joseph~A. Konstan.
\newblock Exploring the filter bubble: The effect of using recommender systems on content diversity.
\newblock In {\em Proceedings of the 23rd International Conference on World Wide Web}, WWW '14, page 677–686, New York, NY, USA, 2014. Association for Computing Machinery.
\newblock \href {https://doi.org/10.1145/2566486.2568012} {\path{doi:10.1145/2566486.2568012}}.

\bibitem{Papadamou2020ItIJ}
Kostantinos Papadamou, Savvas Zannettou, Jeremy Blackburn, Emiliano~De Cristofaro, Gianluca Stringhini, and Michael Sirivianos.
\newblock "it is just a flu": Assessing the effect of watch history on youtube's pseudoscientific video recommendations.
\newblock In {\em International Conference on Web and Social Media}, 2020.

\bibitem{homepage}
Kostantinos Papadamou, Savvas Zannettou, Jeremy Blackburn, Emiliano~De Cristofaro, Gianluca Stringhini, and Michael Sirivianos.
\newblock “it is just a flu”: Assessing the effect of watch history on youtube’s pseudoscientific video recommendations.
\newblock {\em Proceedings of the International AAAI Conference on Web and Social Media}, 16(1):723--734, May 2022.
\newblock URL: \url{https://ojs.aaai.org/index.php/ICWSM/article/view/19329}.

\bibitem{10.1145/3479556}
Kostantinos Papadamou, Savvas Zannettou, Jeremy Blackburn, Emiliano De~Cristofaro, Gianluca Stringhini, and Michael Sirivianos.
\newblock "how over is it?" understanding the incel community on youtube.
\newblock {\em Proc. ACM Hum.-Comput. Interact.}, 5(CSCW2), oct 2021.
\newblock \href {https://doi.org/10.1145/3479556} {\path{doi:10.1145/3479556}}.

\bibitem{filterbubble}
E.~Pariser.
\newblock {\em The Filter Bubble: How the New Personalized Web Is Changing What We Read and How We Think}.
\newblock Penguin Publishing Group, 2011.
\newblock URL: \url{https://books.google.co.ma/books?id=wcalrOI1YbQC}.

\bibitem{patev2021towards}
Alison~J Patev and Kristina~B Hood.
\newblock Towards a better understanding of abortion misinformation in the usa: a review of the literature.
\newblock {\em Culture, Health \& Sexuality}, 23(3):285--300, 2021.

\bibitem{10.5555/3295222.3295319}
Geoff Pleiss, Manish Raghavan, Felix Wu, Jon Kleinberg, and Kilian~Q. Weinberger.
\newblock On fairness and calibration.
\newblock In {\em Proceedings of the 31st International Conference on Neural Information Processing Systems}, NIPS'17, page 5684–5693, Red Hook, NY, USA, 2017. Curran Associates Inc.

\bibitem{10.1145/3289600.3291002}
Bashir Rastegarpanah, Krishna~P. Gummadi, and Mark Crovella.
\newblock Fighting fire with fire: Using antidote data to improve polarization and fairness of recommender systems.
\newblock In {\em Proceedings of the Twelfth ACM International Conference on Web Search and Data Mining}, WSDM '19, page 231–239, New York, NY, USA, 2019. Association for Computing Machinery.
\newblock \href {https://doi.org/10.1145/3289600.3291002} {\path{doi:10.1145/3289600.3291002}}.

\bibitem{beh2}
Manoel~Horta Ribeiro, Raphael Ottoni, Robert West, Virg\'{\i}lio A.~F. Almeida, and Wagner Meira.
\newblock Auditing radicalization pathways on youtube.
\newblock In {\em Proceedings of the 2020 Conference on Fairness, Accountability, and Transparency}, FAT'20, page 131–141, New York, NY, USA, 2020. Association for Computing Machinery.
\newblock \href {https://doi.org/10.1145/3351095.3372879} {\path{doi:10.1145/3351095.3372879}}.

\bibitem{searchbias2}
Bernhard Rieder, Ariadna Matamoros-Fernández, and Òscar Coromina.
\newblock From ranking algorithms to ‘ranking cultures’: Investigating the modulation of visibility in youtube search results.
\newblock {\em Convergence}, 24(1):50--68, 2018.
\newblock \href {https://arxiv.org/abs/https://doi.org/10.1177/1354856517736982} {\path{arXiv:https://doi.org/10.1177/1354856517736982}}, \href {https://doi.org/10.1177/1354856517736982} {\path{doi:10.1177/1354856517736982}}.

\bibitem{youtube70}
Ashley Rodriguez.
\newblock Youtube’s recommendations drive 70\% of what we watch.
\newblock URL: \url{https://qz.com/1178125/youtubes-recommendations-drive-70-of-what-we-watch}.

\bibitem{youtubeRadical}
Kevin Roose.
\newblock The making of a youtube radical.
\newblock URL: \url{https://www.nytimes.com/interactive/2019/06/08/technology/youtube-radical.html}.

\bibitem{misinfo1}
Daniel Röchert, Gautam~Kishore Shahi, German Neubaum, Björn Ross, and Stefan Stieglitz.
\newblock The networked context of covid-19 misinformation: Informational homogeneity on youtube at the beginning of the pandemic.
\newblock {\em Online Social Networks and Media}, 26:100164, 2021.
\newblock URL: \url{https://www.sciencedirect.com/science/article/pii/S246869642100046X}, \href {https://doi.org/10.1016/j.osnem.2021.100164} {\path{doi:10.1016/j.osnem.2021.100164}}.

\bibitem{Sandvig}
Christian Sandvig, Kevin Hamilton, Karrie Karahalios, and C{\'e}dric Langbort.
\newblock Auditing algorithms : Research methods for detecting discrimination on internet platforms.
\newblock 2014.

\bibitem{yttrex}
Leonardo Sanna, Salvatore Romano, Giulia Corona, and Claudio Agosti.
\newblock Yttrex: Crowdsourced analysis of youtube's recommender system during covid-19 pandemic.
\newblock In Juan~Antonio Lossio-Ventura, Jorge~Carlos Valverde-Rebaza, Eduardo D{\'i}az, and Hugo Alatrista-Salas, editors, {\em Information Management and Big Data}, pages 107--121, Cham, 2021. Springer International Publishing.

\bibitem{10.1007/978-3-030-76228-5_8}
Leonardo Sanna, Salvatore Romano, Giulia Corona, and Claudio Agosti.
\newblock Yttrex: Crowdsourced analysis of youtube's recommender system during covid-19 pandemic.
\newblock In Juan~Antonio Lossio-Ventura, Jorge~Carlos Valverde-Rebaza, Eduardo D{\'i}az, and Hugo Alatrista-Salas, editors, {\em Information Management and Big Data}, pages 107--121, Cham, 2021. Springer International Publishing.

\bibitem{sharevski2023abortion}
Filipo Sharevski, Jennifer~Vander Loop, Peter Jachim, Amy Devine, and Emma Pieroni.
\newblock Abortion misinformation on tiktok: Rampant content, lax moderation, and vivid user experiences.
\newblock {\em arXiv preprint arXiv:2301.05128}, 2023.

\bibitem{misinfo2}
Larissa Spinelli and Mark Crovella.
\newblock How youtube leads privacy-seeking users away from reliable information.
\newblock In {\em Adjunct Publication of the 28th ACM Conference on User Modeling, Adaptation and Personalization}, UMAP '20 Adjunct, page 244–251, New York, NY, USA, 2020. Association for Computing Machinery.
\newblock \href {https://doi.org/10.1145/3386392.3399566} {\path{doi:10.1145/3386392.3399566}}.

\bibitem{10.1145/3568392}
Ivan Srba, Robert Moro, Matus Tomlein, Branislav Pecher, Jakub Simko, Elena Stefancova, Michal Kompan, Andrea Hrckova, Juraj Podrouzek, Adrian Gavornik, and Maria Bielikova.
\newblock Auditing youtube’s recommendation algorithm for misinformation filter bubbles.
\newblock {\em ACM Trans. Recomm. Syst.}, 1(1), jan 2023.
\newblock \href {https://doi.org/10.1145/3568392} {\path{doi:10.1145/3568392}}.

\bibitem{sylvia2020we}
Wen-Ying Sylvia~Chou, Anna Gaysynsky, and Joseph~N Cappella.
\newblock Where we go from here: health misinformation on social media, 2020.

\bibitem{youtube4resp}
Youtube Team.
\newblock The four rs of responsibility, part 1: Removing harmful content, 2019.
\newblock URL: \url{https://blog.youtube/inside-youtube/the-four-rs-of-responsibility-remove/}.

\bibitem{NYT}
The New~York Times.
\newblock Tracking the states where abortion is now banned, May 2022.
\newblock URL: \url{https://www.nytimes.com/interactive/2022/us/abortion-laws-roe-v-wade.html}.

\bibitem{homepage2}
Matus Tomlein, Branislav Pecher, Jakub Simko, Ivan Srba, Robert Moro, Elena Stefancova, Michal Kompan, Andrea Hrckova, Juraj Podrouzek, and Maria Bielikova.
\newblock An audit of misinformation filter bubbles on youtube: Bubble bursting and recent behavior changes.
\newblock In {\em Proceedings of the 15th ACM Conference on Recommender Systems}, RecSys '21, page 1–11, New York, NY, USA, 2021. Association for Computing Machinery.
\newblock \href {https://doi.org/10.1145/3460231.3474241} {\path{doi:10.1145/3460231.3474241}}.

\bibitem{rbo}
William Webber, Alistair Moffat, and Justin Zobel.
\newblock A similarity measure for indefinite rankings.
\newblock {\em ACM Transactions on Information Systems}, 28(4):1–38, 2010.
\newblock \href {https://doi.org/10.1145/1852102.1852106} {\path{doi:10.1145/1852102.1852106}}.

\bibitem{10.1145/3485447.3512039}
Maxwell Weinzierl and Sanda Harabagiu.
\newblock Identifying the adoption or rejection of misinformation targeting covid-19 vaccines in twitter discourse.
\newblock In {\em Proceedings of the ACM Web Conference 2022}, WWW '22, page 3196–3205, New York, NY, USA, 2022. Association for Computing Machinery.
\newblock \href {https://doi.org/10.1145/3485447.3512039} {\path{doi:10.1145/3485447.3512039}}.

\bibitem{doi:10.1080/15205436.2020.1714663}
Magdalena Wojcieszak, Stephan Winter, and Xudong Yu.
\newblock Social norms and selectivity: Effects of norms of open-mindedness on content selection and affective polarization.
\newblock {\em Mass Communication and Society}, 23(4):455--483, 2020.
\newblock \href {https://arxiv.org/abs/https://doi.org/10.1080/15205436.2020.1714663} {\path{arXiv:https://doi.org/10.1080/15205436.2020.1714663}}, \href {https://doi.org/10.1080/15205436.2020.1714663} {\path{doi:10.1080/15205436.2020.1714663}}.

\bibitem{wu2023medication}
Jenny Wu, Melissa Greene, Megan Happ, Esm{\'e} Trahair, Melissa Montoya, and Jonas~J Swartz.
\newblock Medication abortion on tiktok: misinformation or reliable resource?
\newblock {\em American Journal of Obstetrics \& Gynecology}, 2023.

\bibitem{10.1145/3373464.3373475}
Liang Wu, Fred Morstatter, Kathleen~M. Carley, and Huan Liu.
\newblock Misinformation in social media: Definition, manipulation, and detection.
\newblock {\em SIGKDD Explor. Newsl.}, 21(2):80–90, nov 2019.
\newblock \href {https://doi.org/10.1145/3373464.3373475} {\path{doi:10.1145/3373464.3373475}}.

\bibitem{tweet}
YouTubeInsider.
\newblock Youtube will remove content that provides promotes false claims about abortion safety, Jul 2022.
\newblock (last accessed December 19, 2022).
\newblock URL: \url{https://twitter.com/youtubeinsider/status/1550153517842587661}.

\bibitem{10.1145/3397271.3401321}
Jiangxing Yu, Hong Zhu, Chih-Yao Chang, Xinhua Feng, Bowen Yuan, Xiuqiang He, and Zhenhua Dong.
\newblock Influence function for unbiased recommendation.
\newblock In {\em Proceedings of the 43rd International ACM SIGIR Conference on Research and Development in Information Retrieval}, SIGIR '20, page 1929–1932, New York, NY, USA, 2020. Association for Computing Machinery.
\newblock \href {https://doi.org/10.1145/3397271.3401321} {\path{doi:10.1145/3397271.3401321}}.

\bibitem{10.1145/3219819.3219886}
Xiangyu Zhao, Liang Zhang, Zhuoye Ding, Long Xia, Jiliang Tang, and Dawei Yin.
\newblock Recommendations with negative feedback via pairwise deep reinforcement learning.
\newblock In {\em Proceedings of the 24th ACM SIGKDD International Conference on Knowledge Discovery \& Data Mining}, KDD '18, page 1040–1048, New York, NY, USA, 2018. Association for Computing Machinery.
\newblock \href {https://doi.org/10.1145/3219819.3219886} {\path{doi:10.1145/3219819.3219886}}.

\bibitem{zimmer2019fake}
Franziska Zimmer, Katrin Scheibe, Mechtild Stock, and Wolfgang~G Stock.
\newblock Fake news in social media: Bad algorithms or biased users?
\newblock {\em Journal of Information Science Theory and Practice}, 7(2):40--53, 2019.

\end{thebibliography}

\end{document}